\documentclass{elsart}

\usepackage{graphicx}

\begin{document}

\begin{frontmatter}



\title{Recoil spectrometers for heavy-ion identification and secondary-beam production: pushing the low-energy limit}


\author[FZK,Orsay]{L.~Audouin\corauthref{thesis}},
\corauth[thesis]{This work forms part of the PhD thesis of L. Audouin}
\ead{audouin@ipno.in2p3.fr}
\author[Orsay]{L.~Tassan-Got},
\author[GSI]{P.~Armbruster},
\author[GSI]{K.-H.~Schmidt},
\author[Orsay]{C.~St\'ephan},
\author[Orsay,GSI,CEA]{J.~Taieb}

\address[FZK]{FZK, Hermann-von-Helmotz Platz 1, 76344 Ettlingen-Leopoldshafen, Germany}
\address[Orsay]{IPN Orsay, CNRS-IN2P3, Campus Paris XI bat. 102, 91406 Orsay, France}
\address[GSI]{GSI, Planckstrasse 1, 64291 Darmstadt, Germany}
\address[CEA]{DEN, CEA Saclay, 91191 Gif-sur-Yvette, France}

\begin{abstract}
The feasibility of low-energy fragmentation experiments using a magnetic spectrometer is discussed. The main challenge is the multiplicity of the ionic charge states, which can hamper the identification in both Z and A of the fragments. Three topics are covered. First, a specific set-up for ionization chambers, based on a very large gas thickness, is presented. Its satisfactory performances are discussed in light of the observations during a 500A~MeV Pb+p experiment performed at the FRS (GSI). As a second topic, the possibility to use a thick layer of matter (a degrader) as a passive measurement device to identify the nuclear charge and the ionic charge state of fragments is discussed. This method, successfully used for Z identification in experiments such as Pb+p at 1A~GeV, fails to measure the charge states at 500A~MeV for the same system. It is shown that surface defects of the degrader are probably responsible for this failure. The third topic is the description of new analysis techniques developed in order to account for and subtract the contribution of polluting charge states in the spectrometer, thus making possible a clean estimation of the production cross sections of all fragments. The combination of those new experimental and analysis techniques made the 500A~MeV spallation experiment a success.
\end{abstract}

\end{frontmatter}

% Shortcuts
\newcommand{\stwo}{$S_2$}
\newcommand{\sfour}{$S_4$}
\newcommand{\mg}{mg.cm$^{-2}$}

%*********************************************************************

\section*{Introduction}

%*********************************************************************

An in-flight fragment separation apparatus can be used in two ways:
\begin{itemize}
\item as a recoil spectrometer, to measure production cross sections or kinematics properties of the fragments;
\item as a separation device, aiming at producing high-purity secondary beams of nuclides.
\end{itemize}

The Fragment Recoil Separator (FRS)~\cite{FRS} of GSI has been very successfully used for both purposes for more than ten years. By now, several projects across the world are dedicated to the construction of more powerful devices, in terms of larger fragment acceptance (spectrometer for fission products), and/or increased selectivity and high-intensity secondary-beam production.

The upper energy range covered by such devices seems to be mostly a technological (and financial) question, as increasing the energy requires more powerful magnets and a longer flight path, not even talking about the accelerator itself. But the lower range is determined by physics: it is assumed that only a very low proportion of non-fully stripped fragments can be tolerated in order to obtain a clean separation of fragments and to measure production cross sections. Therefore, a whole energy range is somehow forbidden to fragmentation experiments.

This low-energy limit has been faced by a recent experiment conducted at the GSI spectrometer FRS. The aim of this experiment was to use the inverse-kinematics method to measure production cross sections from residues of the spallation of lead by protons with an incident energy of 500A~MeV. The results of this experiment will be the subject of a separate, forthcoming publication. This paper is dedicated to the description of the experimental and analysis techniques which were developed for this experiment and to discuss whether or not they proved to be successful. Part of these techniques may be used more generally than for the FRS setup from GSI.

First we will briefly remind the main characteristics of the FRS (section~\ref{chap:frs}). The Fragment Recoil Separator and its set of detectors have been designed using the hypothesis that most, if not all the fragment one wishes to study are fully stripped~\cite{FRS}. This is of course less and less true with decreasing energy. In section~\ref{chap:qst} we will discuss the variations of the charge-state probabilities with energy and their consequences on the selectivity of the spectrometer, as well as the possible choices of so-called stripping materials.

With a lead beam energy as low as 500A~MeV, no stripper is efficient enough so that the fraction of non-fully stripped fragments would be negligible. The response of the ionization chambers (used for the determination of the atomic number of the fragments) in this situation will be discussed in section~\ref{chap:z}, as well as a new setup, developed for this experiment, based on an increased gas thickness.

In section~\ref{chap:deg} we will describe a method to use a thick matter of layer as a passive measurement device in order to estimate the atomic number and the charge state of the fragments. We will also discuss, how unexpected problems prevented the measurement of the ionic charge state of ions (and thus of the mass) in the above-mentioned experiment. The final sections will be dedicated to new analysis techniques aiming at obtaining production cross sections, even in the case of an incomplete event-by-event mass identification. Section~\ref{chap:a} will present a method to determine the most-likely mass of each fragment and to reduce the number of possible contaminating charge states. Then, in section~\ref{chap:cs}, we will explain how to account for the large fraction of non fully-stripped ions and how to obtain isotopic production rates with low uncertainties despite the ambiguities in the mass determination.

We will then conclude and try to answer whether or not the widely admitted low-energy limit has been pushed downwards.

%*********************************************************************

\section{Main characteristics of the Fragment Recoil Separator} \label{chap:frs}

%*********************************************************************

\begin{figure}[ht]
\begin{center}
\includegraphics[width=0.9\textwidth]{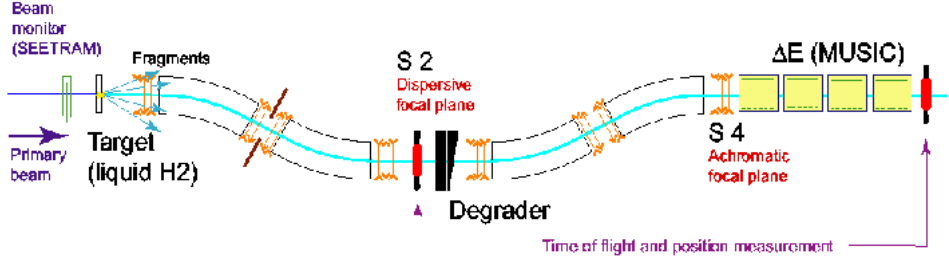}
\caption{\it Schematic view of the Fragment Recoil Separator. Each magnetic section between the focal planes consists of two dipoles (large blocks) plus several quadrupoles and sextupoles (arrows). The horizontal positions at $S_2$ and $S_4$ are usually measured using plastic scintillators, which are also used for time-of-flight measurement.}
\label{fig:FRS}
\end{center}
\end{figure}

The FRS, which is schematically presented in figure~\ref{fig:FRS}, has been designed to separate and identify nuclei produced by fragmentation of relativistic nuclei, the reference beam being 1A~GeV~U~\cite{FRS}. Basically it is designed as a two-stage device: fragments produced in a target located at the entrance of the spectrometer ($S_0$) are selected according to their magnetic rigidity, reach the dispersive focal plane ($S_2$), are then selected again in a second magnetic section and reach the final focal plane ($S_4$). The first magnetic section is intended to measure the magnetic rigidity of the fragments right after the target as well as to remove the primary beam in order to protect the detectors located at $S_2$~and $S_4$.

The behavior of a charged particle of mass $m$ and charge $Q$ moving in a magnetic field $B$ is given by the well-known relation~:
\begin{equation}
B\rho = \frac{m \beta \gamma c}{Q}
\label{eqn:brho_full}
\end{equation}

Here $\rho$ is the curvature radius of the particle trajectory in the magnetic field, $c$ is the speed of light and $\beta$ and $\gamma$ are the usual Lorentz relativistic factors.

For a nucleus with $A$ nucleons, equation~\ref{eqn:brho_full} can be reduced to:
\begin{equation}
B\rho = \frac{u}{e.c} \frac{A}{q} (\beta\gamma)
\label{eqn:brho}
\end{equation}

$u$ being the elementary nuclear mass, $e$ the elementary electric charge, and $q$ the number of ionic charges of the considered ion. In the rest of this paper, for simplicity, we will refer to the magnetic rigidity of a fragment as a dimensionless term, $A/q.\beta\gamma$.

In experiments at the FRS, the $A/q$ ratio of each nucleus is deduced from the measurement of its time of flight and magnetic rigidity in the second magnetic section, using plastic scintillators~\cite{plastic}. $Z$ is deduced from energy loss in ionization chambers set at the exit of the spectrometer.

One of the main hypotheses used for the design of the FRS was that most of the fragments should be fully stripped during their flight in the magnetic sections. In this case one has $q=Z$~and getting the mass of each fragment is straightforward.

As will be outlined in section~\ref{chap:select}, the insertion of a thick layer of matter at $S_2$~is a way to drastically reduce the number of fragments transmitted down to the exit of the spectrometer. Detailed calculations demonstrating this effect can be found in~\cite{degr}. Let us point out that the possibility to choose the exact profile of the degrader allows one to use the FRS in particular magnetic optics modes: achromatic or mono-energetic. \begin{itemize}
\item \it{achromatic}: the position of the fragments at \sfour~does not depend on their velocity at the entrance of the spectrometer. This is particulary helpful to achieve a better separation of the fragments and was the mode selected for the 500A~MeV experiment which is the reference for this paper.
\item \it{mono-energetic}: the velocity of the fragments at \sfour~does not depend on their velocity at the entrance of the spectrometer. This mode is mainly used for implantation experiments.
\end{itemize}

In the following sections, we will often take the $^{208}$Pb+p at 500A~MeV experiment as a reference for our discussions. In this experiment, the target consisted of 87~\mg~liquid hydrogen enclosed in 4 Ti foils, each 9~\mg~thick (see~\cite{target} for a complete description). After this target, a stripper foil of Nb with a thickness of 60~\mg~was set. At \stwo~the plastic scintillator had a thickness of 3~mm and was followed by a profiled degrader with a mean thickness of 1740~\mg~of pure Al. Considering the energy of the fragments after the degrader (roughly 300A~MeV), a Ti foil would have been suitable for optimal stripping, but Al was a good choice, nevertheless (see chapter~\ref{chap:strip} for a discussion about stripper efficiency).

%*********************************************************************

\section{Physics of atomic charge states and consequences} \label{chap:qst}

%*********************************************************************

From equation~\ref{eqn:brho} it is obvious that a magnetic separation device is sensitive to the ionic charge of the nuclei. This is also the case of energy losses (which are the key process for many detectors), as the elemental energy loss in matter is governed by electromagnetic interactions and therefore depends, not on $Z^2$ as one often thinks in the first place, but on $q_{eff}^2$, the square of the effective ionic charge of the nuclei. In the case of a gas medium, this effective charge state is almost equal to the charge state as the time interval between two interactions of the fragment with gas atoms is much larger than the decay time of atomic excited states of the fragment.

\subsection{The physics of atomic charge states: a brief overview}

The evolution of the ionic charge state of an ion in matter depends on two competing processes: electron stripping and electron capture. In the relativistic or quasi-relativistic regime, the stripping probability $\sigma_s$ is nearly independent of the energy of the ion. On the other hand, the probability of electron capture leading to the presence of the electron on shell $X$, $\sigma_c^{(X)}$, strongly increases with decreasing energy (for detailed discussion of the role of the various atomic shells and related phenomena such as the de-excitation time of the capturing ion, see for example \cite{GLOBAL}). One can distinguish two extreme situations:

\begin{itemize}
\item $\sigma_c^{(X)} << \sigma_s$: any electron carried by the ion is removed after a very short distance in the medium. Similarly, any electron capture is quickly followed by the stripping of this electron. Therefore the ion is mostly fully stripped. This situation corresponds to the high-energy limit (roughly, beyond 1A~GeV for heavy ions such as lead).
\item $\sigma_c^{(X)} >> \sigma_s$: any electron stripped for the $X$ shell is quickly replaced. This is most likely to happen if the velocity of the ion is close to the one the electron would have on its shell. In this situation the ion has its $X$ shell filled most of the time. For example, considering heavy ions such as lead, this situation appears in the case of the K shell around 100A~MeV.
\end{itemize}

The intermediate situation ($\sigma_c^{(X)} \simeq \sigma_s$) can be interpreted as a statistical equilibrium between capture and stripping. For a given combination ($Z_{ion}$, $Z_{media}$, $E_{ion}$) one can define a probability $p(q)$ for the ion to have a charge $q$ (or, in other words, to travel with $Z-q$ electrons).

\subsection{Stripping materials}
\label{chap:strip}

Considering equation~\ref{eqn:brho} and the fact that we have no possibility to directly determine the charge state of an ion, it becomes clear that the mass identification of fragments requires, as far as possible, those fragments to be fully stripped in the spectrometer. To optimize the proportion of bare ions, it is convenient to insert layers of well-chosen materials in the device. The choice of a stripper foil results in a compromise between its stripping efficiency and its impact on the beam quality, in terms of angular and energy straggling, and in terms of induced secondary reactions in the stripper.

\begin{figure}[ht]
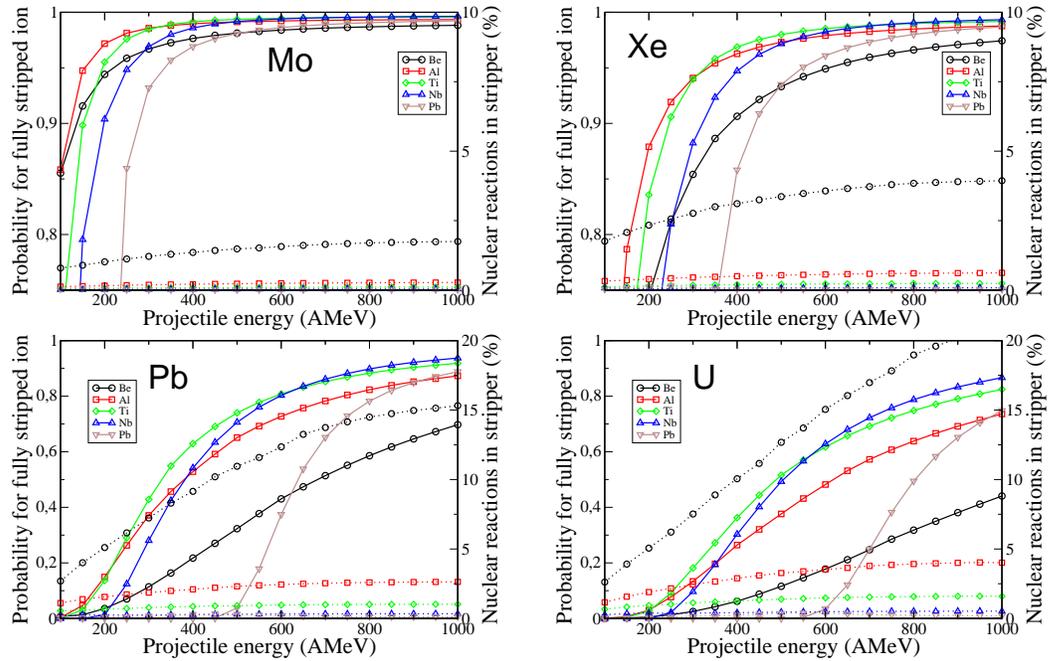

\begin{center}
\includegraphics*[width=0.47\textwidth]{Mo.eps}
\hspace{0.03\textwidth}
\includegraphics*[width=0.47\textwidth]{Xe.eps}
\end{center}
\begin{center}
\includegraphics*[width=0.47\textwidth]{Pb.eps}
\hspace{0.03\textwidth}
\includegraphics*[width=0.47\textwidth]{u.eps}
\caption{\it Bare ion proportion (left scale, continuous lines) and nuclear reaction rate (right scale, dotted lines) in various stripper foils (see legends) for several heavy beams: Mo, Xe, Pb and U, as a function of the incident energy. The thickness chosen for the stripper corresponds to its charge equilibrium thickness at each energy.}
\label{fig:strip}
\end{center}
\end{figure}

Figure~\ref{fig:strip} displays the stripping efficiency, evaluated from the fraction of bare ions, as function of the projectile energy. Calculations were made using the code GLOBAL~\cite{GLOBAL}. The percentage of nuclear reaction in stripper is also displayed (reaction cross sections were estimated using the Karol parametrization~\cite{karol}). In those calculations, to simplify the discussion, the stripper thickness has always been chosen as the so-called charge-equilibrium thickness. This quantity corresponds to the thickness of matter beyond which the proportion of the ionic charge states is fully independent of the initial charge-state distribution.

At high energy, any high- or medium-Z element will provide excellent results. Nevertheless, even at 1A~GeV (the highest energy available at the GSI synchrotron SIS), results are not perfect for very heavy elements: fully stripped ions represent only 95\% of Pb ions, and no more than 88\% for U.

With decreasing energy, the efficiency of strippers is subject to large variations. Nb is the most efficient stripper for beam energies higher than 600A~MeV, while Ti gives slightly better results below this value. Nevertheless, the equilibrium charge state is reached in a thicker foil in Ti than in Nb, which results in a higher nuclear-reaction rate. Let us point out that the energy regime in which the situation can be considered as favorable varies strongly according to the considered ion beam. The 90\% proportion of fully stripped ions, which can be considered as a reasonable limit, is hardly reached for a 1A~GeV U beam, while it allows experiments with a Xe beam at energies as low as 200A~MeV.

\subsection{Impact of charge states on the separator selectivity} \label{chap:select}

The thick degrader technique has been developed in order to allow the production of a high-purity secondary beam. This technique consists in the inclusion of a thick layer of matter between two magnetic selection devices. The first selection basically operates on the $A/q$ ratio of the ions. Those ions loose a large part of their energy in the degrader, which basically depends on the square of their nuclear charge. The second magnetic selection can then be roughly understood as a $Z$ selection.

\begin{figure}[t]
\begin{center}
\includegraphics*[width=0.47\textwidth]{transmitted_stripped.eps}
\hspace{0.03\textwidth}
\includegraphics*[width=0.47\textwidth]{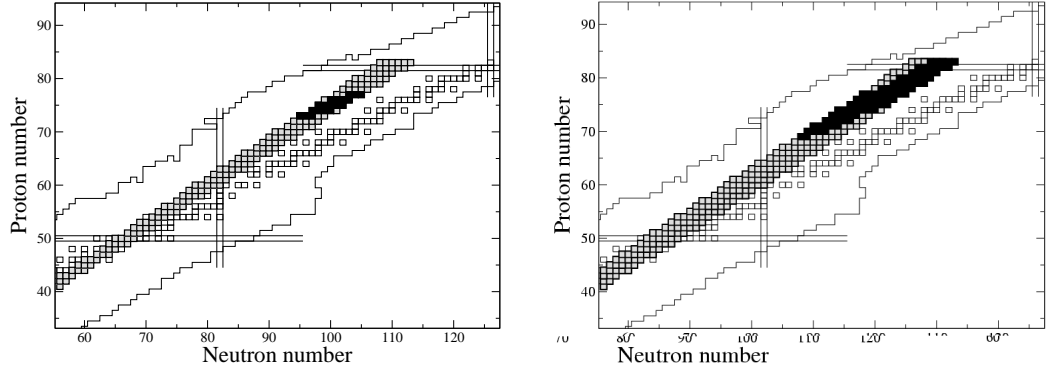}
\caption{\it Simulation of ions transmitted by the FRS spectrometer. The beam is $^{208}$Pb with an energy of 500A~MeV. A thick degrader of thickness 2000~mg.cm$^{-2}$ Al is set between the two magnetic sections. Fragments transmitted in the first part of the spectrometer are plotted in grey, while those also transmitted in the second part are plotted in black. Only fully stripped ions are considered on the left figure, while all combinations with 0 or 1 electron in the first part and 0, 1, 2 or 3 electrons in the second part are considered on the right figure.}
\label{fig:select}
\end{center}
\end{figure}

The selectivity of such a separation device is presented in figure~\ref{fig:select}. Calculations have been performed in order to simulate the transmission in the FRS for fragments produced in the Pb+p reaction at 500A~MeV. We only took into account the fragmentation-evaporation products and did not consider the production cross sections of the reaction products (these cross sections are expected to drop very rapidly for mass losses beyond 40 and completely vanish beyond 60). From the hundreds of different fragments that we considered, less than 20 are transmitted up to the end of the separator. The use of slits in the middle of each magnetic sections would even reduce this number.

Things are less favorable when ionic charge states are taken into account. The various combinations of charge states allow for a much larger population of fragments to be transmitted, as can be seen on the right part of figure~\ref{fig:select}. The purity of a secondary beam is strongly reduced. Consequently, production measurements require more time to achieve satisfactory statistics as a large part of the transmitted fragments may not be of interest.

It is important to notice that, due to the bell shape of the isotopic distributions of evaporation or fission residues, the situation is drastically different if one considers proton-rich or neutron-rich isotopes. If the magnetic device is set to select fully stripped fragments with a given $A/Z$ ratio, H-like fragments with atomic number $Z$ and mass $A-A/Z$ will also be transmitted, as well as He-like fragments of mass $A-2A/Z$, and so on. The proportion of unwanted ions therefore directly depends, not only on the charge-state probability, but also on the production cross sections for the lighter fragments. Let us consider fragments of a given element. In the case of neutron-rich isotopes, cross sections for lighter isotopes are considerably larger, and therefore the proportion of contaminants is expected to be very important. On the other hand, when considering proton-rich fragments, the cross sections for the production of the lightest isotopes are very small, and therefore the problem of charge states may be somehow disregarded.

%*********************************************************************

\section{Z identification} \label{chap:z}

%*********************************************************************

\subsection{Set-up design}

\begin{figure}[ht]
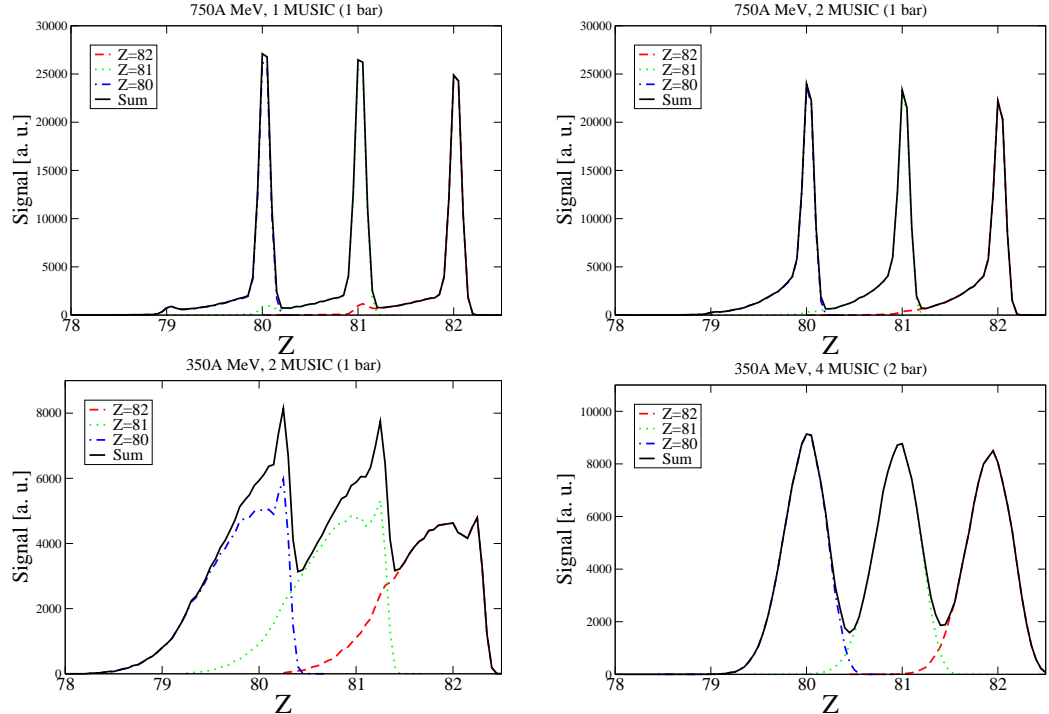

\begin{center}
\includegraphics*[width=0.47\textwidth]{music_simu_750_1.eps}
\hspace{0.03\textwidth}
\includegraphics*[width=0.47\textwidth]{music_simu_750_2.eps}
\end{center}
\begin{center}
\includegraphics*[width=0.47\textwidth]{music_simu_350_2.eps}
\hspace{0.03\textwidth}
\includegraphics*[width=0.47\textwidth]{music_simu_350_8.eps}
\caption{\it Simulation of the energy loss of three fragments (Hg, Tl, Pb) in MUSIC chambers~\cite{MUSIC} (60~cm long each) filled with 1 or 2 bar P10 gas. The fragments were assumed to enter the detectors fully stripped. Incoming energy of fragments is 750A~MeV (upper part) or 350A~MeV (lower part).}
\label{fig:music}
\end{center}
\end{figure}

Calculations presented in figure~\ref{fig:music} summarize the different possible behaviors for stripped ions entering an ionization chamber. The setup in our simulation was intended to simulate the MUSIC chambers~\cite{MUSIC} used at the FRS. In normal conditions, these chambers are filled with P10, a gas mixture with 90\% of Ar and 10\% of CH$_4$, at atmospheric pressure (this corresponds to a gas thickness of 100~\mg). As only 2/3 of these chamber length is instrumented, the ions were propagated along all the gas thickness, but only 2/3 of the thickness was used to calculate the energy collected after deposition in the gas.

The electron stripping and/or capture probability was computed from the cross sections as calculated by GLOBAL~\cite{GLOBAL} for each small step of the fragment propagation through the gas, and the energy loss calculated according to the new ionic charge state of the fragment. Effects such as the escape probability of the $\delta$ electrons or the drift of the electrons through the chambers were not taken into account. The contribution of effects other than the variations of the ionic charge state were roughly taken into account as a fixed 0.3\% contribution to the total straggling.

On the upper part of the figure, the incoming ions have an energy of 750A~MeV and pass through a limited gas thickness, which corresponds to a single ionization chamber filled with 1 bar gas. Their mean free path related to electron capture is much larger than both the gas thickness and the mean free path related to electron stripping. The ions remain stripped along most of their way through the gas, and the achieved resolution in energy loss is excellent. The small tails associated with each peak correspond to the ions that kept an electron for some time in the gas. Of course, adding a second chamber increases the resolving power of the system.

If the incident energy of the fragments is lowered to 350A~MeV, the probability of charge exchanges during the crossing of the fragment through the gas dramatically increases. Therefore, the energy-loss distribution becomes wider as a result of the fluctuations of $q^2$, leading to an unacceptable decrease of the resolution of the ionization chamber. If one increases the total gas thickness (up to 700~mg.cm$^{-2}$ on the figure) the relative statistical fluctuation in $q^2$ is reduced, leading to a better definition of the total energy loss and therefore a better resolution on $q^2$. According to this simulation, a resolution better than 10\% may be expected. Furthermore, complete independency regarding the incoming charge state is achieved.

\subsection{Confrontation with experiment}

According to simulations, the use of ionization chambers to identify heavy, low-energy fragments seems possible if one strongly increases the amount of gas (or the number of chambers) with respect to high-energy setups. This technique has been used successfully during the Pb+p at 500A~MeV experiment at GSI. The setup consisted of 4 MUSIC chambers~\cite{MUSIC} filled with 2 bar P10, for a total gas thickness of the order of 800~mg.cm$^{-2}$, from which roughly 2/3 were instrumented. Due to the presence of a thick degrader at the intermediate focal plane, the energy of fragments in the second part of the spectrometer was of the order of 300A~MeV. The resolution obtained in nuclear charge was better than 0.9\% (or 0.7/Z) for very heavy fragments and better than 0.6\% (or 0.4/Z) for fragments around nuclear charge 70. The resolutions given here are FWHM. We will keep this convention throughout the paper, unless the contrary is specified.

\begin{figure}[t]
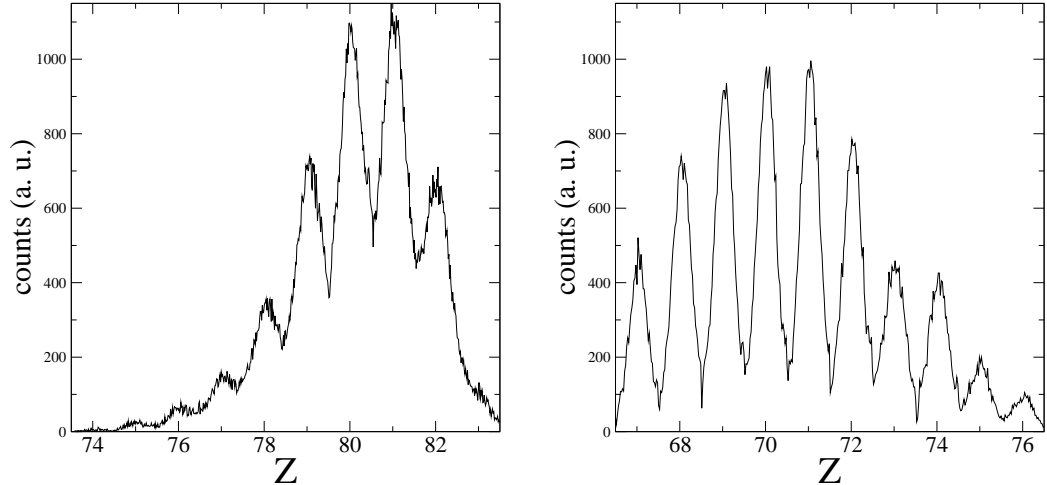

\begin{center}
\includegraphics*[width=0.47\textwidth]{music_185.eps}
\hspace{0.03\textwidth}
\includegraphics*[width=0.47\textwidth]{music_238.eps}
\caption{\it MUSIC spectra obtained in the Pb+p experiment at 500A~MeV for FRS settings centered on $^{183}$Tl (left) and $^{159}$Yb (right).}
\label{fig:music_res}
\end{center}
\end{figure}

Examples of resolutions obtained are presented on figure~\ref{fig:music_res} for fragments close to (left part of the figure) and far from (right part) the lead projectile. This resolution obtained by the combination of the signals of the 4 chambers can be compared with the resolution obtained with a single chamber, which can be deduced from figure~\ref{fig:music_correlation}. To obtain these results, a correction taking into account the velocity of fragments had to be applied.

But the large gas thickness had three drawbacks.

\begin{enumerate}
\item The energy loss is large enough so that a dependency regarding to the mass of the fragment was observed. This is a trivial expectation issued from the Bethe formula, but the consequence is that, for a given $Z$, a systematic increase of the MUSIC signal was observed with decreasing masses. The experimental evidence for this effect is presented in figure~\ref{fig:music_mass}. One can clearly see that this effect is strong enough to lead to a misidentification of the fragments by one charge unit, if not taken into account.

\begin{figure}[t]
\begin{center}
\includegraphics*[width=0.7\textwidth]{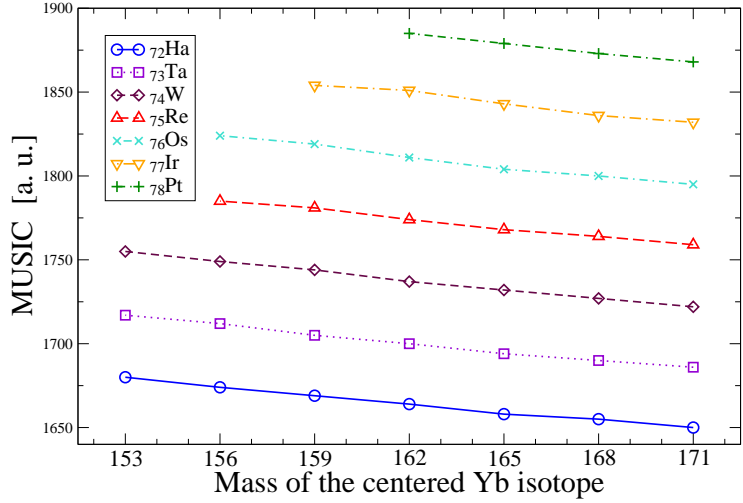}
\caption{\it Mean MUSIC signal for several isotopes in the Pb+p experiment at 500A~MeV. Each abscissa corresponds to a given setting of the magnets of the FRS. In first approximation this defines a cut on the mass-over-charge ratio of the fragments. Considering the limited acceptance in magnetic rigidity of the FRS, each point corresponds to an average over 3 to 5 neighboring masses.}
\label{fig:music_mass}
\end{center}
\end{figure}

\begin{figure}[t]
\begin{center}
\includegraphics[width=0.7\textwidth]{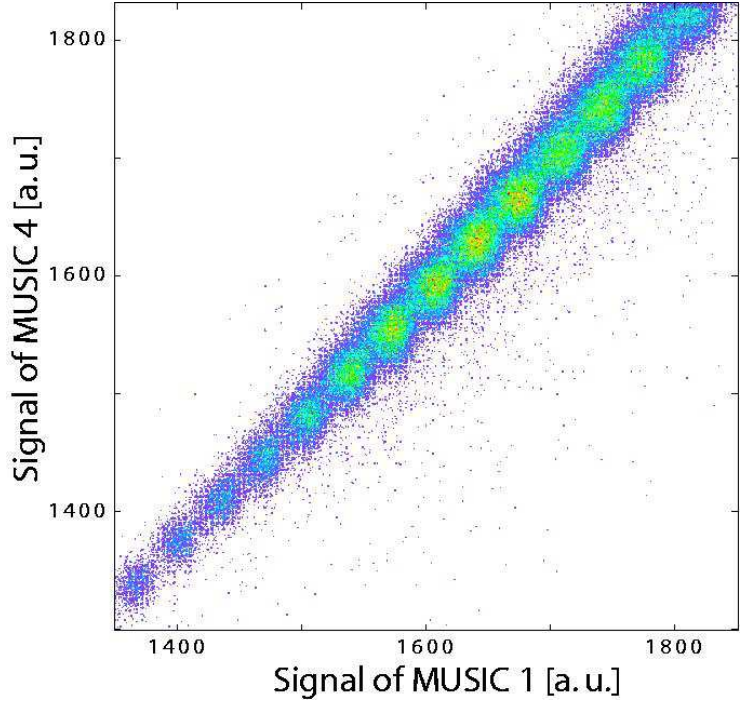}
\caption{\it Correlation of signals from the first and the second MUSIC chambers in a setting centered on $^{159}$Yb in the Pb+p experiment at 500A~MeV.}
\label{fig:music_correlation}
\end{center}
\end{figure}

\item The large amount of matter increases the parasitic nuclear-reaction rate in the gas. Products from these reactions can be rejected during the off-line analysis by comparing the signals observed in the different MUSIC chambers. An example of such a correlation is presented in figure~\ref{fig:music_correlation}.

\item The large quantity of energy deposited in the gas seemed to strongly alter the shape of the relation between the signal of the MUSIC and the position of the fragments in the chambers, as well as the resolution (see section~\ref{chap:countrate}).

\end{enumerate}

\subsection{Effect of large energy deposition in ionization chambers} \label{chap:countrate}

\begin{figure}[ht]
\begin{center}
\includegraphics[width=0.47\textwidth]{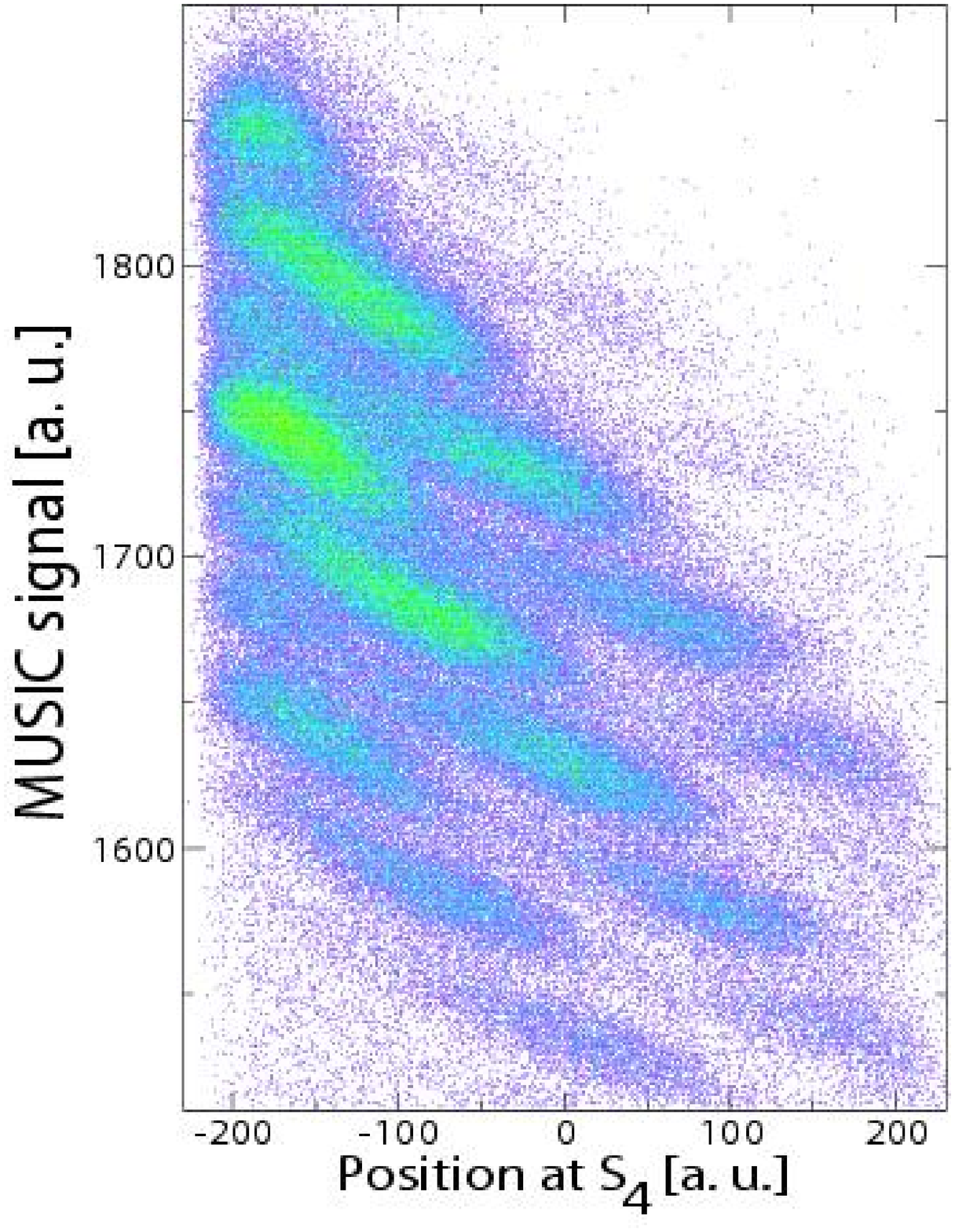}
\hspace{0.03\textwidth}
\includegraphics[width=0.47\textwidth]{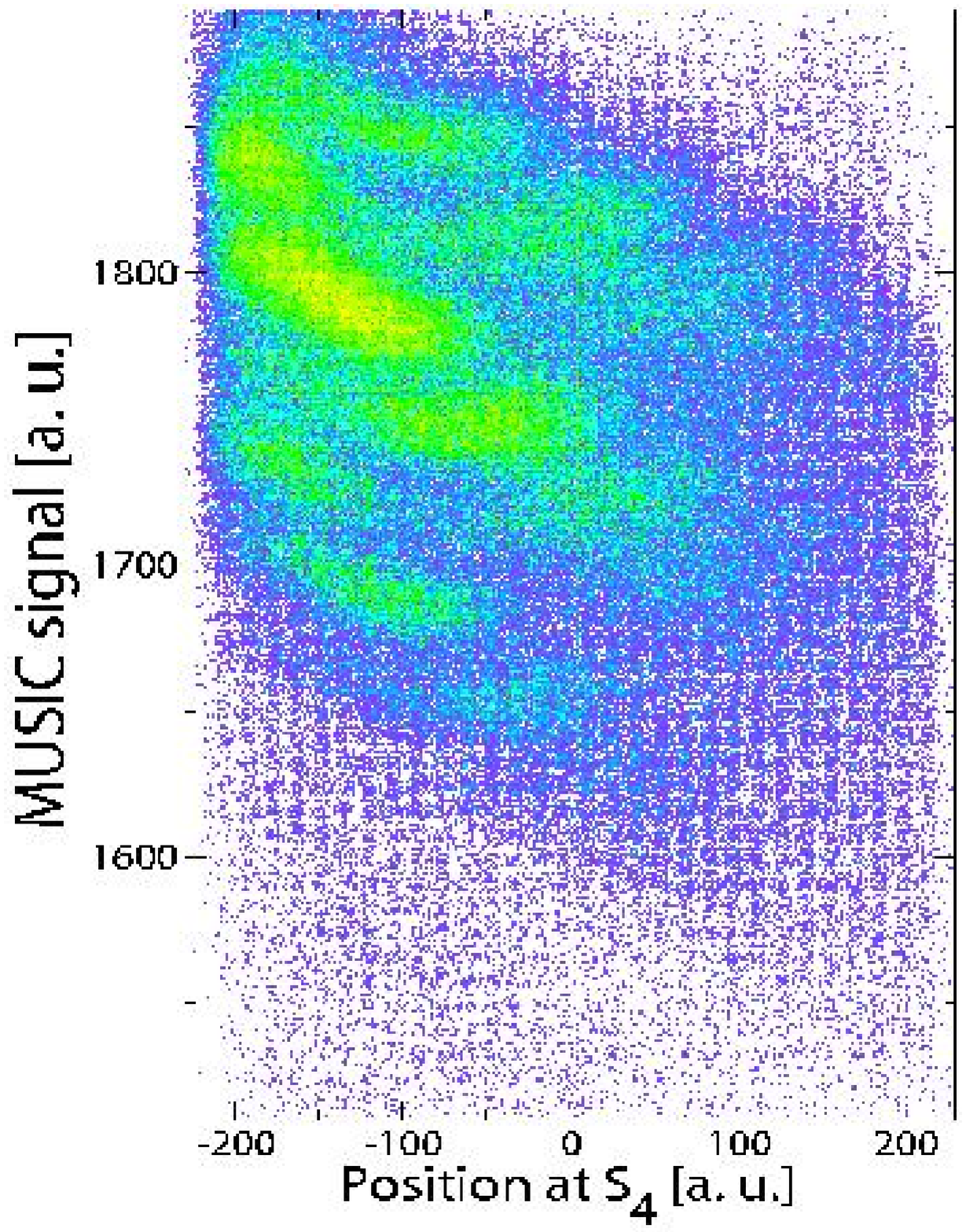}
\caption{\it Sum of the energy deposition in the 4 MUSIC chambers as a function of the distance from the anodes (arbitrary units with zero set in the middle of the chambers) in the Pb+p experiment. Count rate is of the order of 300~Hz (left) and 2~kHz (right).}
\label{fig:rate}
\end{center}
\end{figure}

In ordinary operating conditions, a reduction of the signal is observed with increasing distance between the point of passage of the fragment trajectory and the anodes increases. This dependency is due to the recombination of some of the drift electrons in the gas along their path toward the anodes. Those recombinations may occur with P10 molecules, but are more likely to occur with impurities such as O$_2$ or H$_2$O. The left part of figure~\ref{fig:rate} presents an example of this effect.

If the counting rate is increased up to some 2~kHz, a drastic change of the relation between the signal and the drift distance is observed (right part of the figure~\ref{fig:rate}). Also, compared to the figure on the left, if the resolution is only slightly altered in the region close to the anodes, no identification is anymore possible for trajectories far from the anode.

\begin{figure}[t]
\begin{center}
\includegraphics[width=0.24\textwidth]{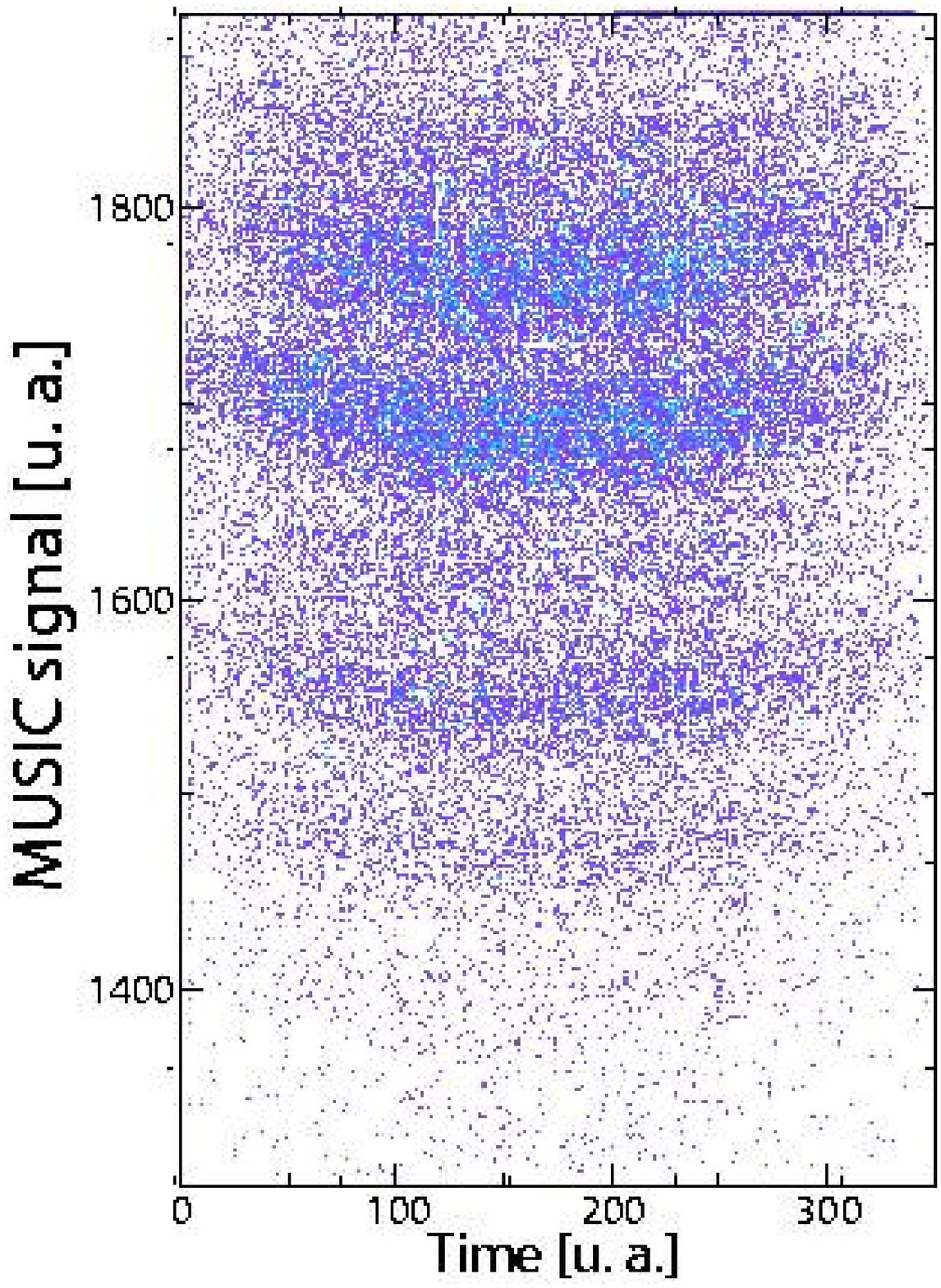}
\includegraphics[width=0.24\textwidth]{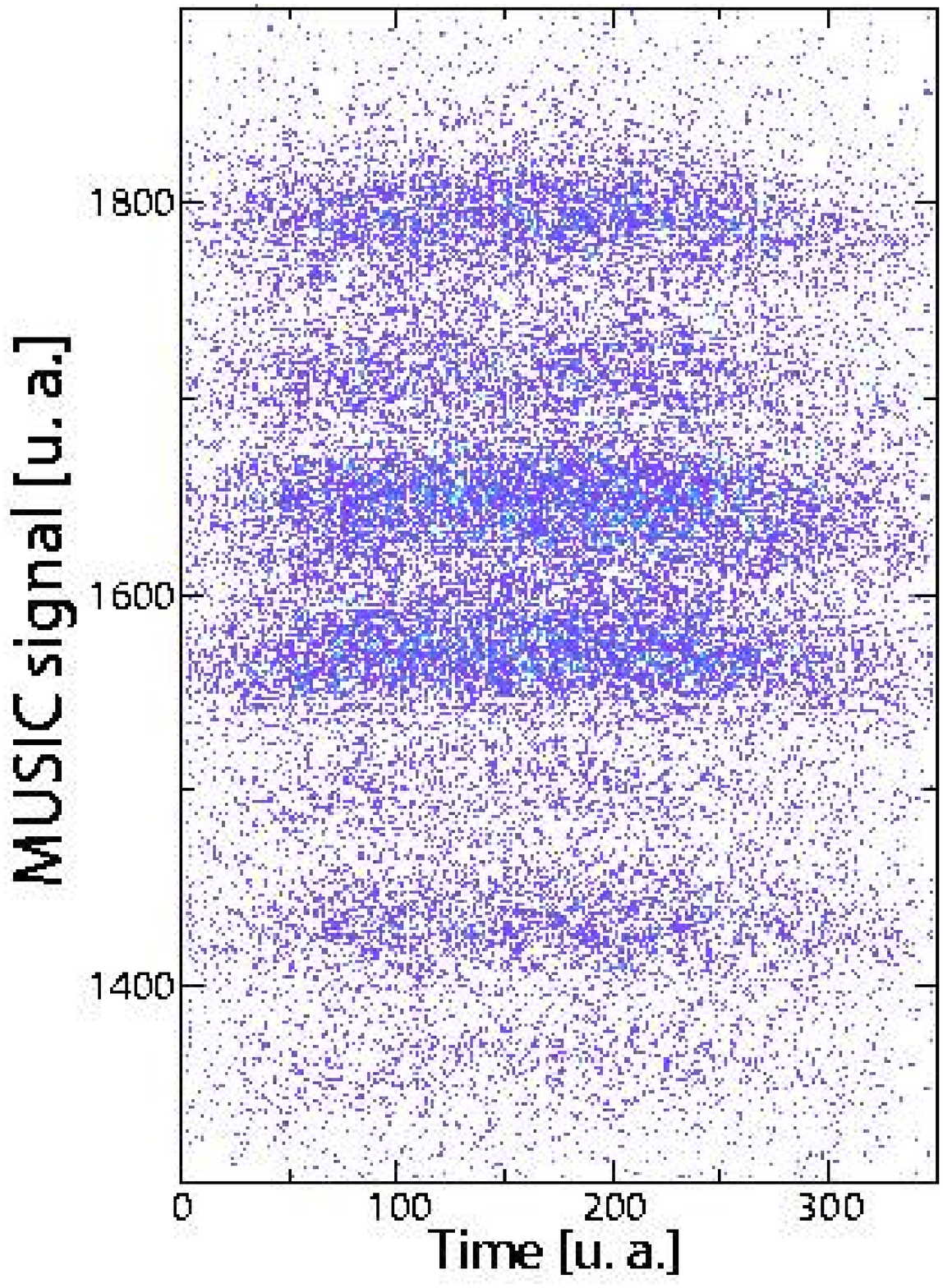}
\includegraphics[width=0.24\textwidth]{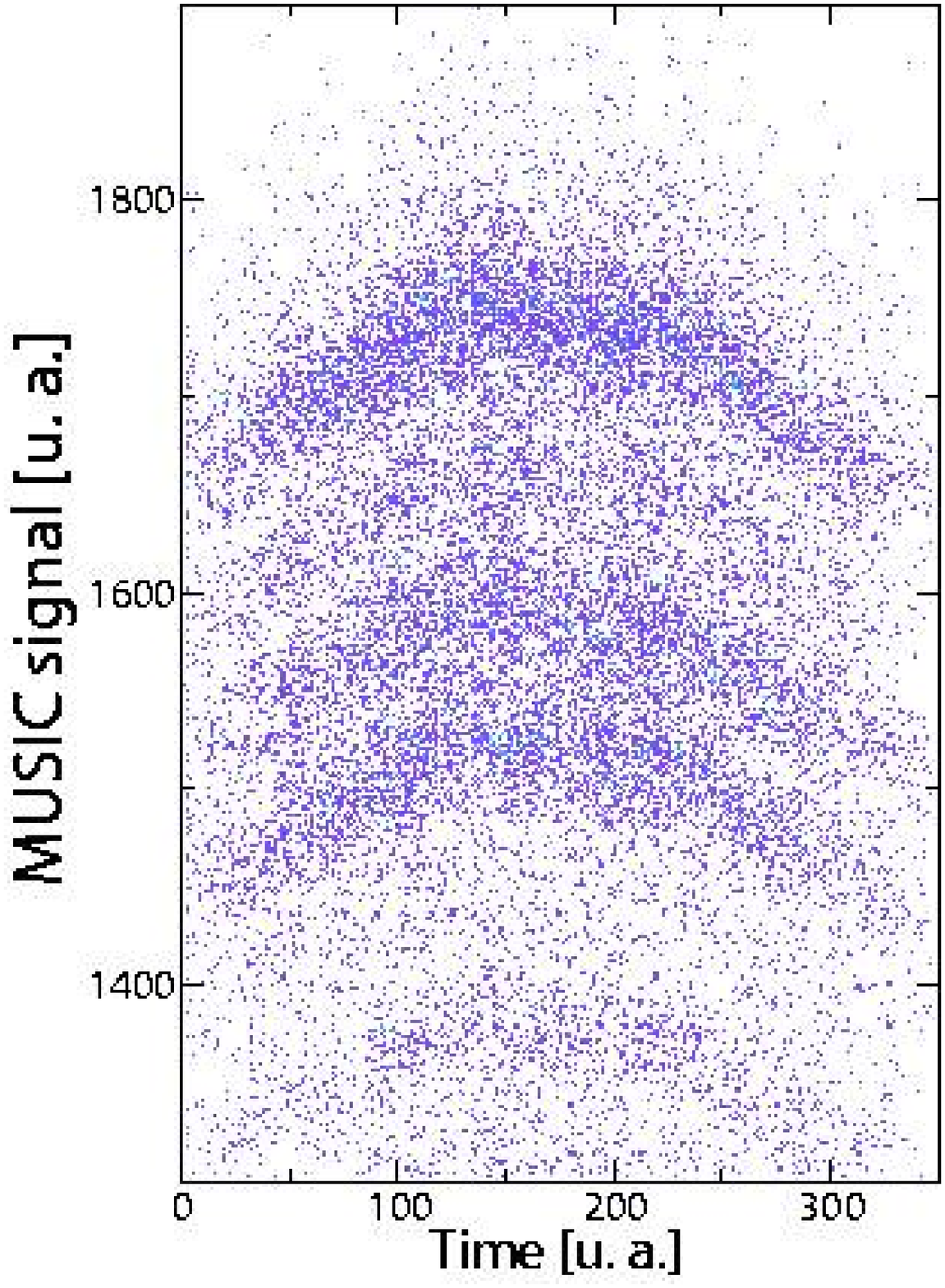}
\includegraphics[width=0.24\textwidth]{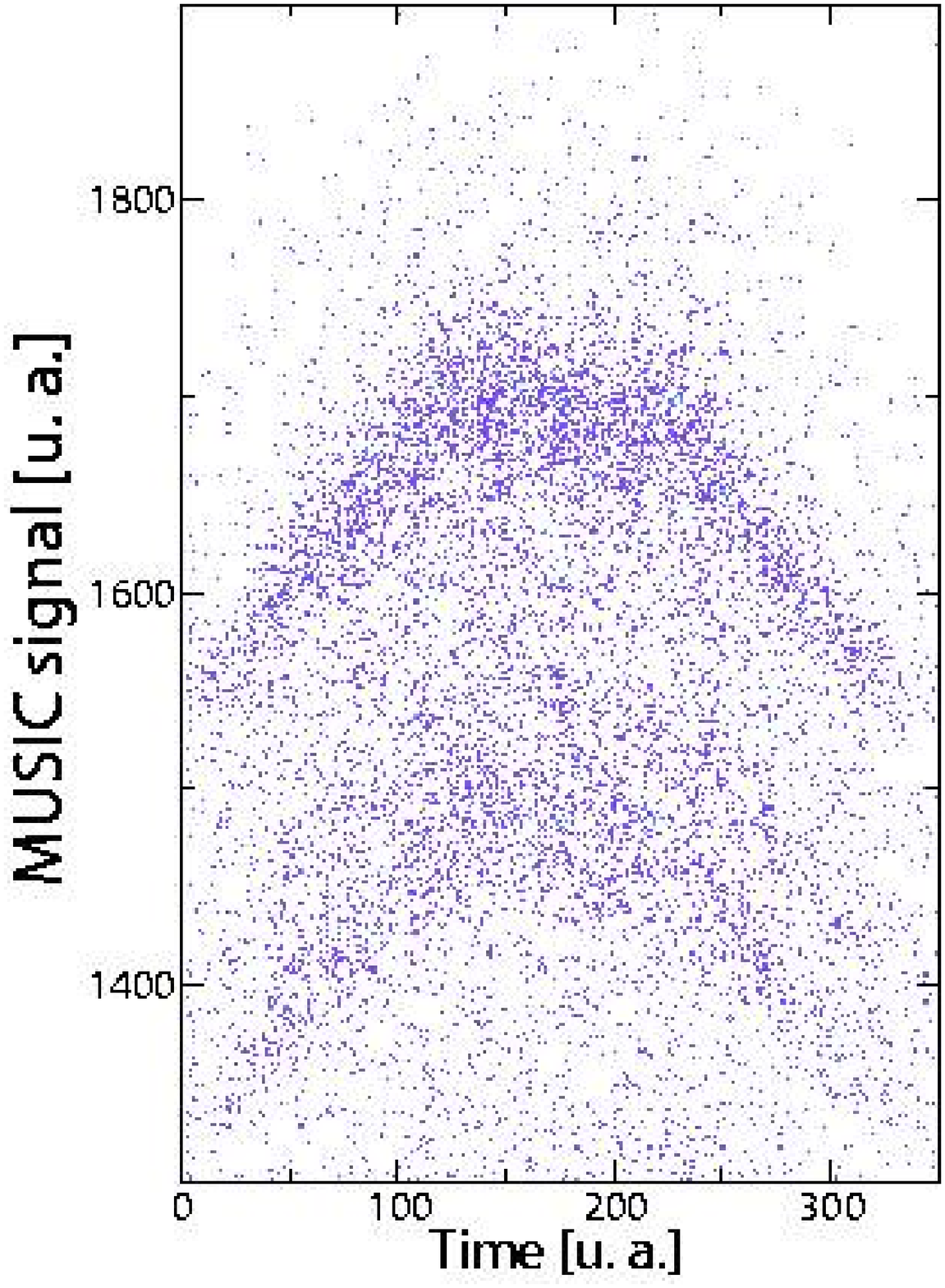}
\caption{\it Sum of the signals of the 4 MUSIC chambers as a function of the time duration of the SIS spills, integrated along different parts of the axis transversal to the trajectory of the fragments. 8 regions of equal length have been defined on this axis. From left to right the figures correspond to the first, third, fifth and seventh region, the first being the closest to the anodes. The counting rate is in the order of 2kHz.}
\label{fig:rate_pos}
\end{center}
\end{figure}

Further illustration of the relation between count rate and resolution in the case of high event counting rates is presented on figure~\ref{fig:rate_pos}, in which the MUSIC signal is plotted as a function of the time duration of a beam spill of the SIS synchrotron. For a section of the MUSIC close to the anodes, the signal slightly reduces with increasing  beam intensity and then gets back to its nominal value at the end of the spill. This limited effect probably is the consequence of the appearance of some screening effect. As one gets further from the anode, the relation inverts: increasing intensity leads to increasing signal, up to a point that leads to a misidentification by 2 nuclear charges. As this effect depends on many parameters it could not be corrected during the analysis, resulting in a loss of some data when the identification of the ions becomes impossible.

As this variation of signal depends on position, electronics can not be responsible for it. This effect is likely to come from the saturation of electron catching sites of the gas (mainly impurities) for high electron densities corresponding to a large number of highly ionizing nuclei.

Such short time variations on MUSIC signals have already been observed in the past without satisfactory explanation. They were tentatively ascribed to pressure or temperature drifts, but no clear correlation with such quantities could be established. We have shown here that this behavior is likely to be related to the electron-attachment saturation under high counting rate, or in other words to the amount of ionization in the chambers.

\subsection{An original method for heavy fragment identification in Z at low energy}

We have described a new way to operate the ionization chambers. We have shown that such a set-up allows to identify in $Z$ heavy ions as heavy as lead at energies as low as 300A~MeV. We are confident that this method would be efficient for lower energies.

The limitations of this set-up have also been stressed. First, the large amount of gas induces variations of the energy loss with the mass of the fragment. This effect does not hamper the efficiency of the method but must be taken into account during the analysis process. Second, as the density of ionization in the gas largely increases with decreasing energy, the response of the ionization chamber can be modified according to the counting rate. This effect can be strong enough to prevent any identification of the fragments. Therefore, it may be necessary to limit the counting rate.

%*********************************************************************

\section{A thick degrader as a passive measurement device} \label{chap:deg}

%*********************************************************************

As written in equation~\ref{eqn:brho}, the mass identification of a fragment requires the knowledge of its ionic charge state. To the best of our knowledge, no apparatus can provide a direct, unambiguous measurement of the ionic charge state of an ion in the energy range considered in this paper. Nevertheless, it may be deduced from a fine evaluation of the energy loss in a thick layer of matter, namely the degrader set between the magnetic sections of the FRS.

In this section, we will consider that the fragments are correctly identified in $Z$, regardless of their charge state, according for example to the use of the method presented in the previous section. For the sake of simplicity, we will often call "degrader" the whole matter inserted in the beam line at \stwo, namely, the degrader plus the plastic scintillator. They were of course treated separately in the simulation.

\subsection{Energy loss in the degrader}

The energy loss in matter of an ion of mass $A$ and kinetic energy $E$ is given by:
\begin{equation}
\Delta E = A \Delta \gamma
\label{eqn:dE_0}
\end{equation}

For calculations related to magnetic rigidities, it is more convenient to write equation~\ref{eqn:dE_0} as a function of $\Delta (\beta\gamma)$ instead of $\Delta \gamma$. This leads to:
\begin{equation}
\Delta E \simeq A \Delta (\beta\gamma) <\beta>
\label{eqn:dE}
\end{equation}
where $<\beta>$ is the mean value of $\beta$ in the considered layer of matter. In the energy domain we are discussing in this paper, the variations of this quantity are small and one can therefore treat it as a constant for the purpose of the demonstration.

The slowing down of the fragments and the change of charge state (if any) induce a variation of the magnetic rigidity between magnetic sections. Using the discussion that followed equation~\ref{eqn:brho}, this can be written as:

\begin{eqnarray}
\Delta (B\rho) & = & \frac{A}{q_1} (\beta\gamma)_1
                   - \frac{A}{q_2} (\beta\gamma)_2 \\
               & = & \frac{A}{q_2} \left( (\beta\gamma)_1 - (\beta\gamma)_2 \right)
                   - \left( \frac{A}{q_2} (\beta\gamma)_1 - \frac{A}{q_1} (\beta\gamma)_1 \right)
\end{eqnarray}

Schematically the first $A/q_2$ term (which we will note as $\Delta (B\rho) ^{(E)}$) corresponds to the loss of energy, while the second term (which we will note as $\Delta (B\rho) ^{(q)}$) is related to the charge-state changing.

According to equation \ref{eqn:dE} one can write:
\begin{equation}
\Delta (B\rho) ^{(E)} = \frac{\Delta E}{m_0c^2q_2}\frac{1}{F}
\end{equation}

According to the discussion in section~\ref{chap:qst}, $\Delta E$ can be written:
\begin{equation}
\Delta E = q_{eff}^2 f(\beta) g(X_{S_2})
\end{equation}
where $q_{eff}^2$ is the mean value of the square of the atomic charge of the fragment in the degrader (as defined in chapter~\ref{chap:qst}), $f(\beta)$ is the velocity dependence from the Bethe formula, and $g(X_{S_2})$ describes the position dependence due to the variations of the degrader thickness.

According to its definition, $q_{eff}^2$ is completely independent of the ionic charge state in both parts of the spectrometer. Furthermore, it is both unambiguously related to $Z^2$ and very close to it. We can then replace it by $Z^2$ for this calculation. We can now write $\Delta (B\rho)^{(E)}$ as:
\begin{equation}
\Delta (B\rho) ^{(E)} = \frac{Z^2 f(\beta) g(X_{S_2})}{q_2}\frac{1}{F} \propto \frac{Z^2}{q_2} \simeq Z + e_2
\label{eqn:dbrho1}
\end{equation}

Here $e_2$ stands for the number of electrons in the second section of the spectrometer. The second part of the development of $\Delta (B\rho)$ writes:
\begin{equation}
\Delta (B\rho) ^{(q)} = (B\rho)_1.\frac{q_2-q_1}{q_2}
\label{eqn:dbrho2}
\end{equation}

To summarize, we have established that, besides the initial velocity of the fragment, the variation of magnetic rigidity depends on three physical components:
\begin{itemize}
\item the nuclear charge of the fragment;
\item the charge state of the fragment in the second part of the spectrometer;
\item the change of charge state of the fragment between the two sections of the spectrometer.
\end{itemize}

In the next section (\ref{chap:de_app}) we will present the possible applications of the measurement of magnetic-rigidity variation and the associated requirements on the energy-loss resolution. In section~\ref{chap:de_res} we will discuss the limitations from the characteristics of the devices and from physics. In section~\ref{chap:de_exp} we will present and discuss the results obtained in the 500A~MeV experiment.

\subsection{Applications of the measurement of the energy loss in the degrader}
\label{chap:de_app}

Among the three effects listed in the previous section, the charge-state changing is always dominant as soon as a capture or a ionization takes place in the degrader, even with a very thick degrader for which the energy loss is large. This allows to detect and identify unambiguously any change of charge state as will be shown later in figure~\ref{fig:deg}. The related magnetic-rigidity variation does not depend on the thickness of matter set, and therefore information on a change of charge state is accessible whatever the setup or the energy of the incoming fragment. During the data analysis, the fragments can be sorted according to their change of charge state, a quantity that we will write as $\Delta q$.

Besides this property, and as expected from equation~\ref{eqn:dbrho1}, a large thickness of matter induces a linear dependence of the variation of magnetic rigidity with $Z$ and $q$. Separating peaks associated to neighboring charges with a reasonable efficiency requires the FWHM of each peak to be (roughly) smaller than 0.7/$Z$. For example in the case of lead this corresponds to a resolution of 0.9\%.

In high-energy experiments, the selection of the charge states of the fragment is nearly entirely obtained using the $\Delta q$ value. Indeed, although the proportion of fully stripped ions (that we can write (0,0) according to the number of electrons they carry in each part of the spectrometer) is only of the order of 80\%, the $\Delta q$ estimation allows the rejection of most of the charge-state combinations like (0,1), (0,2) or (1,0). The only charge-state combinations which can not be disentangled from (0,0) are (1,1), (2,2), and so on. But in the case of the fragmentation of a 1A~GeV Pb beam for example, these charge states represent only 2\% of the Pb fragments that have a $\Delta q$ equal to zero. This very low proportion is due to the combination of the two strippers (one at the entrance of each magnetic section) which is weighted by the product of the two probabilities. These fragments might be disregarded during the identification process and receive an appropriate correction during the estimation of production rates. In this case it is therefore convenient to use the degrader as an additional measurement of the $Z$ value, which can be combined to the one obtained from the MUSICs and increase the overall $Z$ resolution. This method has already been successfully used in previous spallation experiments at the FRS~\cite{Au_Fanny,Pb_p,Pb_d,U_Taieb,U_Enrique}.

In the case of a low-energy experiment, the fraction of $\Delta q = 0$ fragments can be as low as 40\% for fragments produced from a 500A~MeV Pb beam. Among those $\Delta q = 0$ fragments, roughly 40\% are hydrogen-like ions. Therefore, they should be explicitly identified and taken into consideration. If one relies on the MUSICs for the measurement of the nuclear charge, according to the previous calculations, a precise estimation of the energy loss in the degrader may be used to estimate the ionic charge state of the fragments.

\subsection{Resolution limitations imposed by physics and by detectors}
\label{chap:de_res}

Let us first stress what exactly means the word resolution. The value we are considering is the variation of the magnetic rigidity. It is important to distinguish the two sources of the so-called uncertainty on this quantity. The first and most obvious one is the uncertainty on the measurement. The second one, by far the largest as we will see, is not an uncertainty in the usual sense of the word: it is the fluctuation in energy induced by the path through a thick layer of matter. Whatever the precision of the measurement of the magnetic rigidity or time of flight, there is no way to correct the $\Delta B\rho$ width induced by those fluctuations. Therefore, to identify charge states, it is the sum of those two terms which must fulfill the 0.7/$Z$ condition presented above.

For convenience, one can express the straggling in energy as a fluctuation of the variation of the magnetic rigidity:

\begin{equation}
\frac{ \delta ( \Delta E )}{\Delta E} \simeq \frac{\delta(B\rho)}{(B\rho)_1 - (B\rho)_2}
\end{equation}

This relation is discussed in details in appendix~\ref{chap:brho-e}. In the following we will express all fluctuations and uncertainties as relative uncertainties with respect to the variation of magnetic rigidity.

We will now discuss separately the contributions to the total $\Delta B\rho$ width of the energy straggling and of the magnetic-rigidity measurement.

\subsubsection{Measurement of the magnetic rigidity and associated uncertainty}
\label{chap:uncer}

The resolution that can be achieved on the measurement of the change of magnetic rigidity obviously depends on the resolution in magnetic rigidity, which itself depends on the precision of the magnets and from the position-measurement devices, namely scintillators.

In a given magnetic setting between the focal planes $A$ and $B$, characterized by a magnetic rigidity $(B\rho)_0$ for centered ions, the magnetic rigidity of an ion is deduced from its position at the focal planes (respectively $x_A$ and $x_B$) as:

\begin{equation}
\label{eqn:xbrho}
(B\rho) = (B\rho)_0 . \left( 1 + \frac{x_B}{D} - M\frac{x_A}{D} \right)
\end{equation}
with $D$ being the dispersion of the magnetic section and $M$ its magnification.

An optimum result is obtained only if the magnetic optics exhibits no aberration and if the positions of the fragments are measured at the focal planes of each magnetic section. We will assume that the first condition is fulfilled, although there are indications that some limited magnetic aberrations may exist close to the final focal plane.

For the second assumption, the situation differs according to the focal plane one considers. In the first part of the FRS, the initial point (the target) is unambiguously defined, and the scintillator is set at the intermediate focal plane, at a fixed position. The magnification term in equation~\ref{eqn:xbrho} vanishes and the uncertainty is simply obtained from the uncertainty on magnetic fields (known to be less than 10$^{-4}$) and on the scintillator response (estimated to be at most 3~mm):

\begin{equation}
\frac{\delta(B\rho)_1}{(B\rho)_1}
= \frac{\delta B_1}{B_1} + \frac{\delta x_2}{D}
\simeq 3.10^{-4}
\end{equation}

The position of the final focal plane varies according to the chosen magnetic optics, and the position of the corresponding scintillator depends on the available room left by all other detectors located at this region. Therefore, the two points may not coincide. If the distance between the scintillator and the focal plane is $d$ and the angular distribution of a fragment has a width $\sigma_{\theta}$, a dispersion is introduced on the measurement of the $x_4$ position:

\begin{equation}
\delta x_4 = \sigma_{\theta}.d
\end{equation}

The contribution of this term to the uncertainty on the measured magnetic rigidity adds to the intrinsic uncertainty of the measurement:

\begin{equation}
\left( \frac{\delta(B\rho)_2}{(B\rho)_2} \right)^2 \simeq
( 5.10^{-4} )^2 + (\sigma_{\theta}.d)^2
\end{equation}

If the trajectory angle of the fragments is measured, the position of fragments at the final focal plane can be reconstructed. In this case, the enlargement on the position resolution due to the incoming angle can therefore be corrected.

\subsubsection{Energy straggling}
\label{chap:e_strag}

Two effects are responsible for the fluctuations of the energy loss of a fragment passing through a solid~\cite{Geissel1,Geissel2}. First, the energy loss is due to successive collisions, each removing a variable amount of energy from the incoming ion. The variance can be estimated with a reasonable precision by:

\begin{equation}
\frac{d\sigma^2(dE)}{dx} = 4\pi \left( \frac{e^2}{4\pi \epsilon_0} \right)^2 N_A Z_m Z_f^2 \gamma^2 \chi
\end{equation}
where $Z_m$ is the nuclear charge of the material, $Z_f$ the nuclear charge of the fragment passing through, and $N_A$~the number of atoms of the material. This term will decrease the resolution as the thickness of the material increases and as the energy of the fragment increases. In the Bohr approximation $\chi$ reduces to $1-\beta^2/2$. The Lindhard-Sorensen (LS) theory~\cite{LS} has extended the validity of this equation to relativistic energies. In this theory $\chi$ is a much more complicated expression but the agreement with experiment in the relativistic regime is strongly improved. For our subsequent calculations of energy losses and straggling, we used this theory as implemented in the code ATIMA~\cite{ATIMA}.

The second phenomenon inducing straggling is the fluctuation of the ionic charge state, already discussed in section~\ref{chap:z}. These fluctuations are of statistical nature, therefore the energy loss resolution increases with the thickness of the solid. On the other hand, fluctuations increase with decreasing energy as more and more charge states can be populated. The calculations related to this phenomenon were performed using the ionic capture and stripping cross sections from the code GLOBAL~\cite{GLOBAL}.

\begin{figure}[ht]
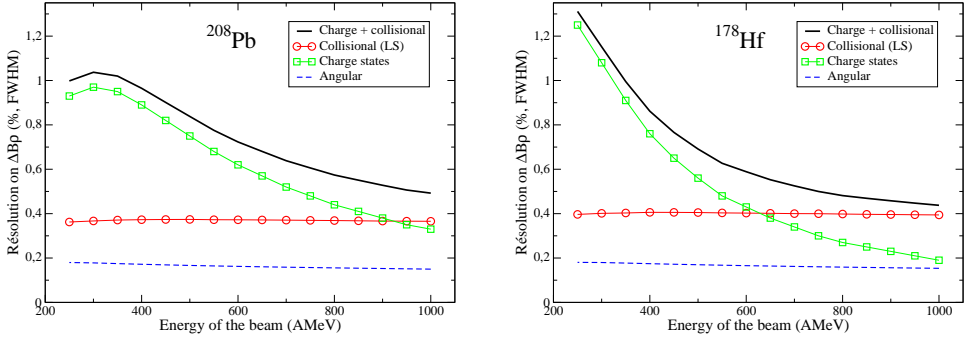

\begin{center}
\includegraphics[width=0.43\textwidth]{Pb_strag.eps}
\hspace{0.5cm}
\includegraphics[width=0.43\textwidth]{Hf_strag.eps}
\caption{\it Calculations of the straggling on the energy loss (FWHM) due to collision straggling (circles) and charge state fluctuations (squares) for $^{208}$Pb (left plot) and $^{178}$Hf (right plot), as the function of the energy of the beam. The continuous line represents the quadratic sum of these two components. Effect of the angular straggling is presented in the hypothesis of a distance of 1~m between the position measurement at the exit of the spectrometer and the position of the magnetic focal plane. See text for details of the simulation.}
\label{fig:sigma_de}
\end{center}
\end{figure}

We have performed calculations of these two effects for ions of $^{208}$Pb (left part of figure~\ref{fig:sigma_de}) and $^{178}$Hf (right part). For each energy we calculated the energy loss of a $^{208}$Pb beam in a liquid hydrogen target used in all spallation experiments performed at the FRS (see chapter~\ref{chap:frs}), but did not consider any straggling or angular broadening at this point. Any energy straggling in the target is supposed to be washed out by the achromatism of the system. This is not the case of the angular distribution induced by the reaction, but we disregarded it in order to focus on the effects of the matter set at \stwo. The thickness of the degrader was then calculated for each energy as it is done during experiments: the total thickness of matter at the intermediate focal plane must correspond to half of the range of the beam in aluminium (this value represents an optimum between the energy loss and the probability for parasitic nuclear reactions in these layers). We explicitly took into account a plastic degrader made of 3~mm of BC-400.

Results of these calculations are expressed in terms of FWHM for the relative magnetic rigidity variation. At 1A~GeV the charge-state fluctuations are comparable to the effect of the collision straggling in the case of Pb, while, due to the quadratic summation, they are nearly negligible in the case of Hf. The total FWHM for Pb ions is 0.49\%, and 0.44\% for Hf.

With decreasing energy, the collision straggling keeps a nearly constant value, while the change state changes intervene more and more. At 500A~MeV, the resolution on energy loss for lead is not better than 0.84\%, the collision straggling accounting for only 20\% of this value.

Unexpectedly, in the case of Hf, the worsening of the resolution with decreasing energy is stronger than in Pb and does not reach a maximum as it does for Pb. This is probably related to the number of collisions in the materials, which increases faster for Pb than for Hf with decreasing energy, therefore reducing the statistical fluctuations on the charge state. Furthermore, around 250A~MeV, the cross section for capture to the $s$~atomic shell of Pb ions is already very large while the capture cross section to the $l$-shell remains small: this strongly favors the $q=Z-2$~charge state and thus furthers reduces the fluctuations. In the case of Hf this situation appears only at an energy lower than 200~MeV.

We did not extend the calculations to lower energies as the energy loss in the target would have become too large to render an experiment feasible. Reducing the hydrogen thickness would not counteract this problem in a relevant way as it would also lead to an increase of the relative contribution of the target windows (which cannot be reduced) to the production rates.

We also calculated the effect of the angular straggling induced by the materials set at the intermediate focal plane. In chapter~\ref{chap:uncer} we have already discussed this effect: in an achromatic system like the FRS, it varies linearly with the distance between the real magnetic focal plane and the point of the beam line at which the position measurement is performed (it is therefore canceled if the position is measured at the correct point of the beam line). The values plotted in figure~\ref{fig:sigma_de} correspond to a distance of 1~m between the expected and the actual measurement points. As one can see on the figure such a misplacement has nearly no consequence on the obtained energy loss resolution. Of course a larger distance could play a significant role.

\subsubsection{Experimental beam width at the final focal plane}

\begin{figure}[t]
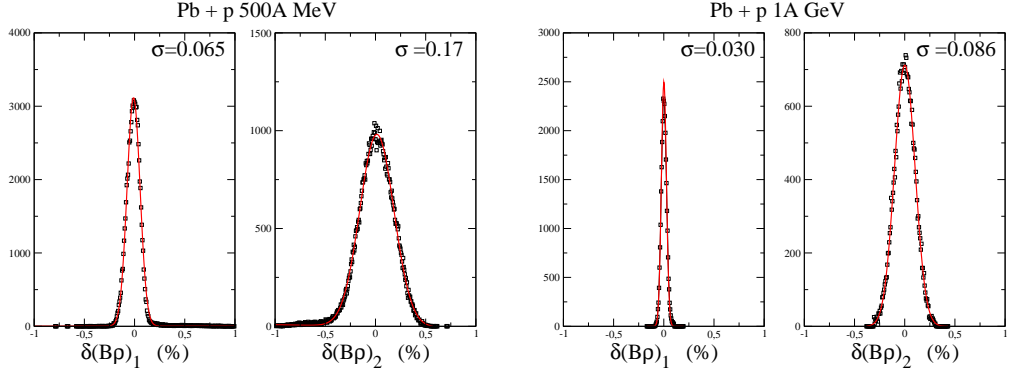

\begin{center}
\includegraphics*[width=0.45\textwidth]{beam_500.eps}
\hspace{0.5cm}
\includegraphics*[width=0.45\textwidth]{beam_1000.eps}
\caption{\it Beam width in percent of deviation of magnetic rigidity, as deduced from the positions measured at \stwo~and \sfour~in the Pb+p experiments at 500A~MeV (left) and 1A~GeV (right). Squares correspond to experimental points, and a Gaussian fit has been performed on each spectrum (solid curve).}
\label{fig:beam_pos}
\end{center}
\end{figure}

As we have already discussed, in an achromatic system, the position of an ion at the final focal plane does not depend of its initial velocity, nor of its initial angle. Going through the degrader induces velocity fluctuations that can not be compensated by the magnetic optics. Therefore, the beam is not expected to be perfectly focused at \sfour. The previous calculations on the energy straggling can be converted in a minimal width expected for the beam at \sfour: the results are 0.27\% at 500A~MeV and 0.16\% at 1A~GeV. The experimental values are presented in figure~\ref{fig:beam_pos}. Beam widths at \stwo~are also plotted for comparison ; let us recall that the dispersion observed at the intermediate focal plane is expected to be fully corrected by the achromaticity of the system.

In the case of the 1A~GeV experiment, the width observed at \stwo~corresponds to a magnetic-rigidity straggling of 0.06\% (FWHM relative to B$\rho_1$). An enlargement is observed at \sfour, leading to a magnetic-rigidity straggling of 0.2\% (FWHM relative to B$\rho_2$, thus 0.54\% relative to $\Delta$B$\rho$). This is in good agreement with the result of the calculation presented above (0.49\% FWHM relative to $\Delta$B$\rho$).

In the case of the 500A~MeV experiment, the width observed at \stwo~corresponds to a magnetic-rigidity straggling of 0.16\% (FWHM relative to B$\rho_1$). A very large value is observed at \sfour, corresponding to a magnetic rigidity straggling of 0.4\% (FWHM relative to B$\rho_2$, thus 1.24\% relative to $\Delta$B$\rho$). This is much higher than the result of the calculation presented above (0.26\% FWHM relative to B$\rho_2$). Let us consider the different possible reasons for this large position spread.

\begin{itemize}

\item The absence of correction for the angle of the ions. In this discussion we only consider the beam and therefore this possibility can be ruled out. Even if the scintillator is far from the magnetic focal plane (2~m can be considered as an upper boundary), the position broadening induced by the angle is not expected to be a major contributor to the overall broadening (the angular aperture of the beam has been estimated by our calculations to be of the order of 2~mrad at this point of the beam line).

\item The energy straggling and the charge state fluctuations. This obviously plays a role: we have shown in section~\ref{chap:e_strag}, that precise calculations lead to a value of 0.26\% (FWHM relative to B$\rho_2$) for this effect. This induces an important broadening of the position spectra (which is not compensated by the achromaticity of the system) but is not sufficient to explain the observed value.

\item A failure in the design of the degrader leading to a chromaticity of the system. If one rules out an error in the calculation of the degrader profile, the hypothesis of surface defects or inhomogeneities remains.

\end{itemize}

From the difference between the calculated widths at \sfour~at the two considered energies, we can postulate a typical size of the inhomogeneities. We could reproduce the observed widths at both energies by considering that the degrader surface is inhomogeneous with very small defects: 27~$\mu$m. This value is to be compared to the degrader thicknesses: roughly 6.5~mm in the 500A~MeV experiment and 15~mm at 1A~GeV. The relative straggling in magnetic rigidity difference induced solely by those inhomogeneities would be 1\% at 500A~MeV, and only 0.3\% at 1A~GeV.

We consider the presence of such inhomogeneities plausible as their effect would be small enough not to affect experiments conducted at higher energies, which necessarily include thicker degraders. Therefore they were never considered as potentially harmful for experiments.

\subsection{Obtained resolution on the variation of magnetic rigidity in high- and low-energy experiments}
\label{chap:de_exp}

\begin{figure}[ht]
\begin{center}
\includegraphics*[width=0.7\textwidth]{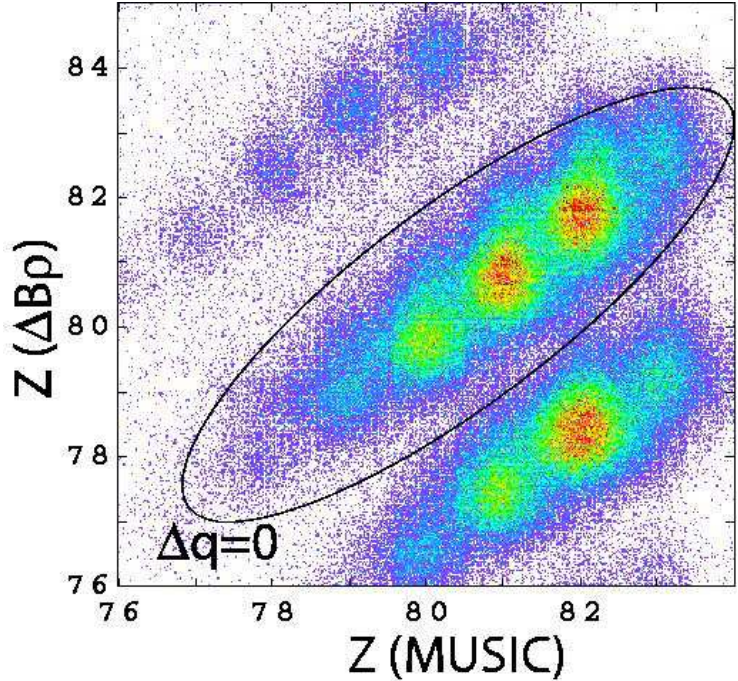}
\caption{\it Correlation between the variation of magnetic rigidity (normalized to the nuclear charge for fully stripped fragments) and the $Z$ value obtained from MUSICs in the Pb+p at 500A~MeV experiment. The central band corresponds to fragments that did not gain or loose any electron while passing through the degrader. The upper band corresponds to fragments that lost an electron, the lower band to fragments that gained one electron.}
\label{fig:deg}
\end{center}
\end{figure}

The use of a thick degrader as a passive measurement device has already proved to be successful in previous experiments conducted at the FRS at energies of or close to 1A~GeV. We have already emphasized that its use in the spallation of lead at 500A~MeV was of particular importance as the number of non-stripped ions was expected to be very high (up to 60\% after the degrader).

At higher energies, the influence of degrader inhomogeneities become negligible, and the charge-state straggling is strongly reduced. Our calculations indicate that a 0.5\% resolution should be obtained, which allows for a satisfactory $Z$ or $q$ identification. This conclusion is in good agreement with results from previous experiments.

From figure~\ref{fig:deg}, it is obvious that during the 500A~MeV experiment the resolution required to separate charge states was not achieved: no separation can be seen between fully stripped and H-like fragments. They appear as a bump, clearly visible for each spot in the upper part of the most intense line, but impossible to disentangle from the fully stripped fragments. This could be expected from the discussion conducted in the previous section: the position straggling observed at \sfour~leads to an uncertainty of 0.4\% on the magnetic rigidity. As the difference of magnetic rigidity is only 31\% of the magnetic rigidity in the second part of the spectrometer, the uncertainty on $\Delta B\rho$ is expected to be close to 1.25\%, which is very close to the observed resolution. Such a value prevent identification from the various charge states of the fragments. In the discussion above we only considered the beam, which allowed us to neglect the influence of the angle on the precision of the magnetic-rigidity variation. This is not true anymore when one looks at the fragments, and a non-optimal position of the scintillator at \sfour~may have further worsen the experimental resolution. We could not put in evidence any effect of the angle in the case of the 500A~MeV experiment as a failure of a device prevented the measurement of the angle of the fragments at \sfour. This does not affect our global conclusion on the reasons why the reduced resolution was obtained for the energy loss measurement in the degrader.

Based on our simulations and the experimental results, we postulate that defects on the degrader surface are the main reason for this very poor resolution, and that an improved degrader (with a surface roughness smaller than 10~$\mu$m) may allow this method to be successfully used at an energy as low as 500A~MeV for very heavy beams such as lead.

%*************************************************************************

\section{Evaluation of fragment masses} \label{chap:a}

%*************************************************************************

The determination of the mass of each fragment requires two steps, which will be described in this section. First, the A/q ratio is deduced from the magnetic rigidity and time of flight in the second section. Then, if the use of the degrader has been successful in determining the ionic charge state of the fragments, or if the proportion of non-stripped ions is low enough to neglect them, getting the mass of each fragment is straightforward.

Alternatively, if the charge state is not known, it is convenient to determine a "most-likely" mass for each fragment.

\subsection{A/q ratio}

The $A/q$ ratio is deduced from the magnetic rigidity and from the time of flight in the second section of the spectrometer. Various corrections of dependencies from the positions have to be applied: effect of the position of the fragment at \stwo~and \sfour~on the length of the trajectory, non-linear effects in the time response of the scintillators. This leads to a parametrization defined as:

\begin{equation}
\label{eqn:aq}
A/q = f(x_2,x_4,ToF) = \frac{L_0 + L_2 x_2 + L_4 x_4}{(T_0 - ToF) + T_2 x_2^2 + T_4 x_4^2}
\end{equation}

$L_i$ and $T_i$ are free parameters. They are assumed to stay constant as long as magnetic optics is not changed. Their values can be adjusted considering a given setting for which masses are affected by counting, starting from the centered fragment. It is usually convenient to use a magnetic setting corresponding to fragments close to the beam, as identification is in this case quite straightforward because of the reduced velocity spread of those fragments. In contrast, in low-energy experiments the high proportion of charge states suggests a setting centered on very light isotopes. The calculated parameters can be checked by re-calculating the mass number of the ion beam. In the case of the 500A~MeV experiment, the mass of $^{208}Pb$ was obtained with a precision of $5.10^{-4}$.

\subsection{A/q resolution and effect of the polluting charge states}

The resolution required to separate masses with a peak-over-valley ratio of 10 is 0.1\%. The FRS layout and detectors have been designed to fulfill this requirement. In this respect, the most difficult situation is expected to be the measurement of heavy fragments at high energy: in this case the time of flight is short and the difference in magnetic rigidity between two neighboring masses is minimum. In this respect, a low-energy experiment is expected to be a better case, as the first constraint is softened.

Nevertheless, the presence of polluting charge states can strongly hamper the resolution obtained on the A/q ratio. For fragments produced from a heavy, stable nucleus such as lead, the A/q ratio is close to 2.5. Therefore the A/q peaks for fragments with an odd number of electrons will fall in-between peaks associated to fragments with an even number of electrons, leading to an apparent loss of resolution. This has of course no consequence if the ionic charge state of each fragment is known, as the exact mass can be obtained. On the other hand, if the ionic charge state is not known, one can assign its most-likely mass to each fragment. Also, according to the previous discussion, the fragments of mass $A$ are mixed with fragments $A-2$ and $A-3$. We will show that one can use the correlation between the $A/q$ ratio and the $\Delta B\rho$ values to reduce the number of actual masses that are mixed together.

%The obtained $A/q$ resolution varies from 0.1 to 0.4\%, which is not satisfactory as  This apparent lack of resolution is indeed a proof of the reliability of the method, as the bottom of valleys correspond to H-like ions for which $A/q$ ratio is shifted of roughly 0.5 units due to the $A/Z$ ratio of fragments considered in this experiment (which is close to 2.5). An improvement of the overall resolution is clearly needed. This can be achieved using the correlation between the energy loss in the degrader and the calculated $A/q$ ratio.

\begin{figure}[ht]
\begin{center}
\includegraphics[width=0.45\textwidth]{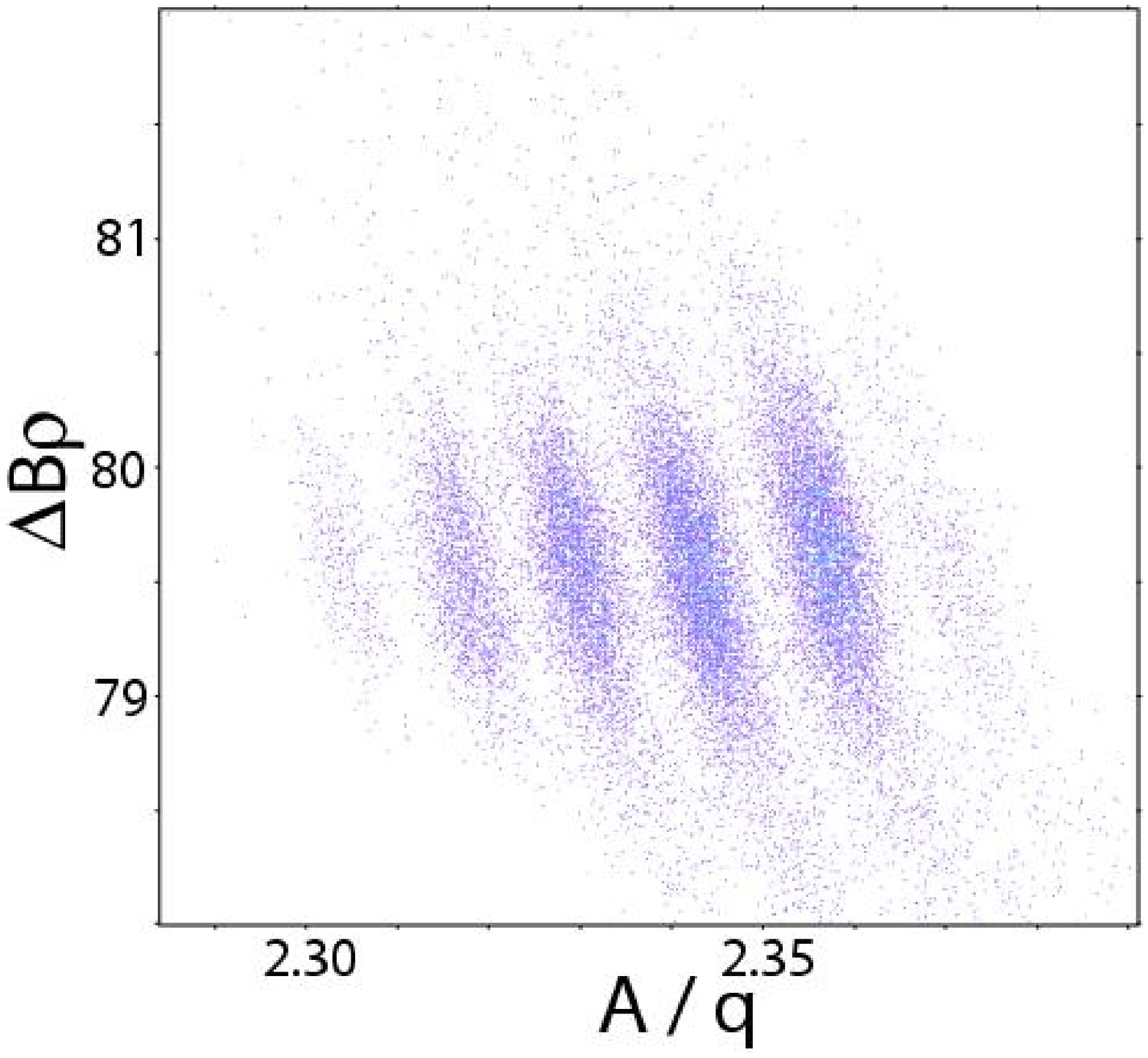}
\hspace{0.05\textwidth}
\includegraphics[width=0.45\textwidth]{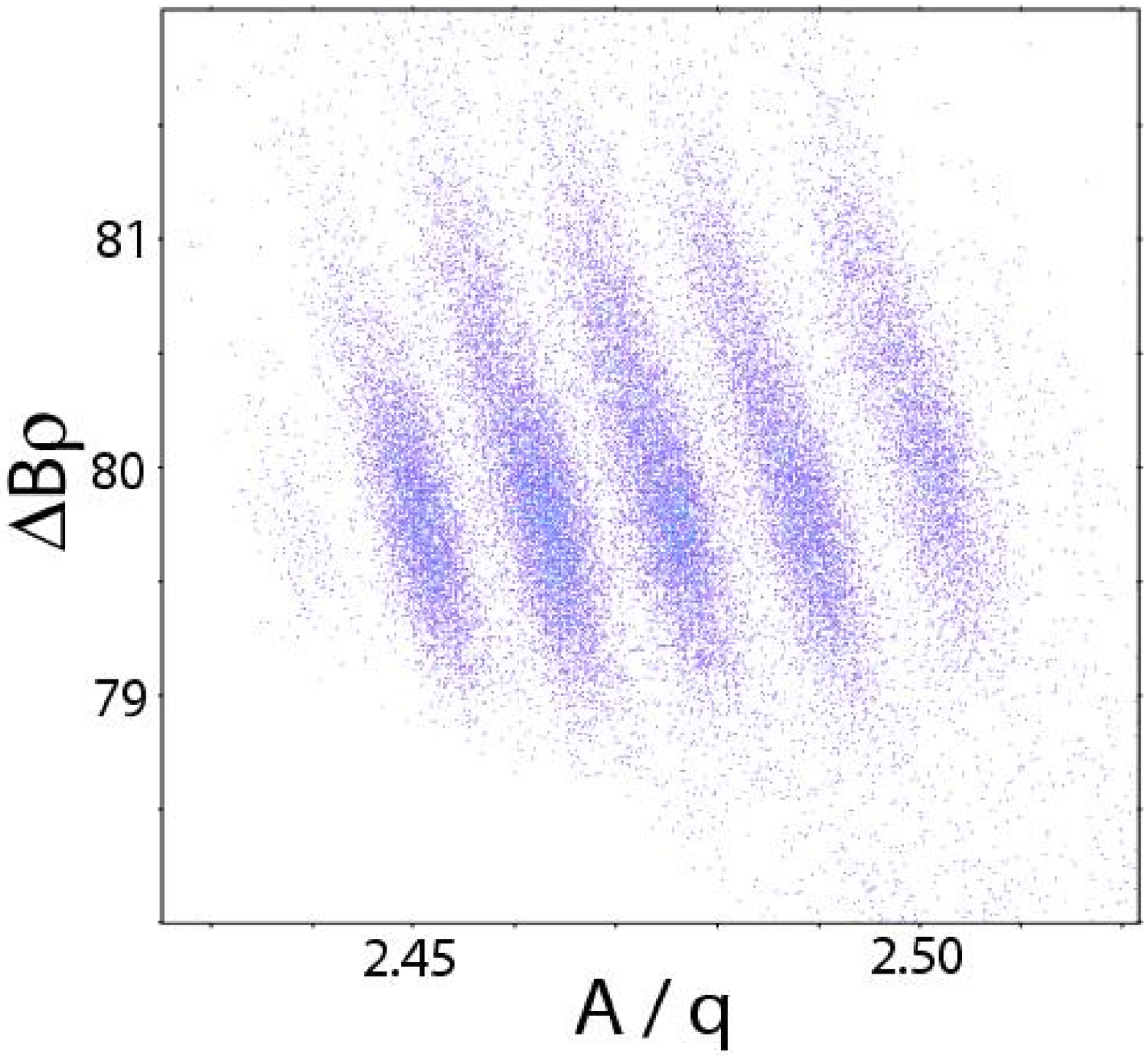}
\caption{\it Correlation between the differences of magnetic rigidity and the calculated A/q ratio for Hg ions. Left figure corresponds to a FRS setting centered on $^{180}Hg$, right part to a setting centered on $^{198}Hg$.}
\label{fig:aq_dbrho}
\end{center}
\end{figure}

An example of such a correlation is presented in figure ~\ref{fig:aq_dbrho}, for heavy and light Hg isotopes produced in the 500A~MeV Pb+p experiment (only fragments with $\Delta q = 0$ are represented). This figure also underlines the lack of resolution in $\Delta B\rho$ obtained in this experiment. The large bands seen for heavy isotopes are a mix of the fully stripped ions (low part of bands) and H-like ions (upper part). In the case of light isotopes, very few H-like ions are produced and transmitted, and therefore only the lower part of the bands remains.

As no direct separation of charge states is possible, it is advantageous to get the $A/q$ value, not from a direct projection on the $A/q$ axis, but from a projection along the direction of the correlation bands. Not only does it improve the apparent resolution, creating clearly separated peaks (in the case of the 500A~MeV experiment, the resolution becomes  close to 0.1\% for all settings, while it could reach 0.4\% in the most unfavorable cases). It also ensures that only masses $A-3$ contribute to the pollution of isotopes of mass $A$. The unavoidable weakness of this method is that the group of fragments identified as being of mass $A$ and atomic charge $q_0$ contains a variable fraction of $(A-3,q=q_0-1)$ fragments, which will have to be subtracted in order to deduce the production rates.

At this step unambiguous identification in mass on an event-by-event basis is of course not possible anymore. Nevertheless it is convenient to assign a mass to each fragment. To calculate masses, we assume that fragments gaining zero or more electrons when passing through the degrader were fully stripped in the first part of the FRS, while fragments loosing one or more electrons are fully stripped in the second part of the FRS. In other terms, for a given value of $\Delta q$, each fragment is assumed to be in the most probable charge state in the two parts of the FRS. In the worst case (heavy fragments for which $\Delta q = 0$) this hypothesis is valid for only 60\% of fragments.

%*********************************************************************

\section{Production rates for ambiguously identified fragments} \label{chap:cs}

%*********************************************************************

Even if fragments are not unambiguously identified in mass on an event-by-event basis, we will show that it remains possible to extract isotopic production rates. This requires to evaluate and correct for the fraction of fragments that was not correctly identified because of the unresolved ionic charge states. In this chapter we will present a method to achieve those corrections.

\subsection{Kinematic spectra}

First of all it is necessary to take into account the limited acceptance of the FRS. Its acceptance is limited to $\pm$1.5\% in magnetic rigidity (or in longitudinal momentum) for each given ion. This means that, if the broadening of the fragment velocity goes beyond this value, only part of the momentum distribution will be transmitted in a given setting of magnets. Calculation of the production cross sections therefore requires the reconstruction of the complete kinematic distribution of each fragment, which may require data from different magnetic settings.

The momentum from each fragment in the first part of the spectrometer can be calculated by applying the relation (\ref{eqn:brho}) to the first part of the FRS:
\begin{equation}
(\beta \gamma)_1 = \frac{(B\rho)_1}{A/q_1}
\end{equation}

For identified fragments, $A$ and $q$ are known as integer values. Therefore, the precision obtained on the velocity depends only on the precision on the magnetic rigidity. As previously discussed, it can be estimated to be in the order of $3.10^{-4}$, which is far better than the best time-of-flight measurements for heavy ions.

\begin{figure}[ht]
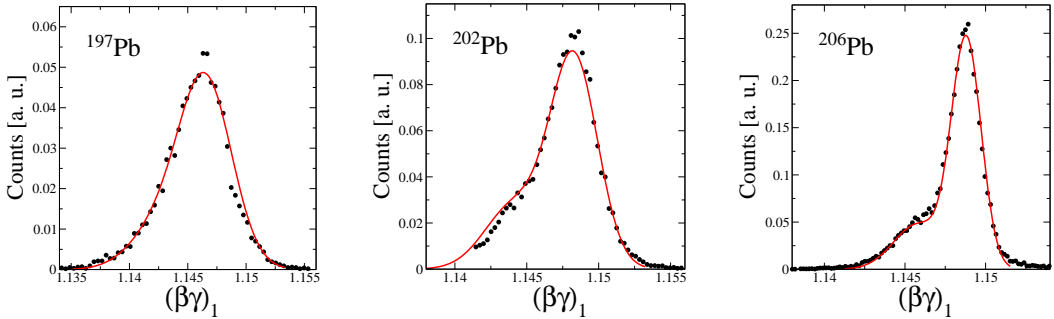

\begin{center}
\includegraphics[width=0.3\textwidth]{kin_197.eps}
\hspace{0.03\textwidth}
\includegraphics[width=0.3\textwidth]{kin_202.eps}
\hspace{0.03\textwidth}
\includegraphics[width=0.3\textwidth]{kin_206.eps}
\caption{\it Velocity distribution for lead isotopes in the first part of the FRS during the 500A~MeV Pb+p experiment. Only fragments that kept the same charge state in both parts of the spectrometer are represented. The continuous lines correspond to the fit described in the text.}
\label{fig:bg}
\end{center}
\end{figure}

Any error on the estimation of the charge state of fragments has a direct consequence on the calculated mass and therefore on the calculated momentum value. Let us consider for example a fragment $(Z,A)$ which was assumed to be fully stripped in both parts of the FRS. If this fragment was in fact an hydrogen-like ion, its real mass is $A-3$ and its calculated momentum is related to its real one by:

\begin{eqnarray}
(\beta \gamma)_1^{(calculated)} & = & \frac{(B\rho)_1}{(A-3)/Z-1} \;.\; \frac{A-3}{A} \frac{Z}{Z-1} \\
& = & (\beta \gamma)_1^{(real)} \;.\; \frac{A-3}{A} \frac{Z}{Z-1}
\end{eqnarray}

The velocity distributions of fragmentation residues are known to be of Gaussian shape. As it can be seen on figure~\ref{fig:bg}, the misidentified fragments create a second component in the velocity spectra. According to the previous discussion about the fragmentation experiments, this component is very important for high-Z, neutron-rich fragments while it slowly vanishes for low-Z and/or proton-rich fragments.

There are two ways to deal with this contribution: either a direct estimation through a fit, or an estimation of the whole distribution (also by a fit) followed by a subtraction of the part contributed by the pollution. If the first solution is well suited for fragments very close to the projectile, for which the velocity spread is small enough for the two peaks to be distinguished, its application to the case of lighter fragments, for which the two peaks partially or totally overlap, would lead to large uncertainties. Therefore, we recommend the second solution, and we will now present a way to implement it in a numerically stable way.

The velocity distribution of each fragment can be described as the sum of two Gaussian functions, the second one accounting for the pollution related to misidentified fragments. In order to optimize its stability, a single fit is applied, for each set of data associated to a given combination of $Z$ and $\Delta q$. The coefficients describing the center (noted $ (\beta \gamma)_0$) and the width (noted $\sigma$) of each Gaussian are related to one another by simple, physics-based relations derived from the Morrissey systematics~\cite{morrissey}:

\begin{equation}
(\beta \gamma)_0^{(Z)}(A) = \alpha_0(Z) + \alpha_1(Z).A
\end{equation}
\begin{equation}
\sigma^{(Z)}(A) = \sqrt{\beta_0(Z) + \beta_1(Z).A + \beta_2(Z).A^2 }
\end{equation}

Defining an elementary velocity distribution as:

\begin{equation}
G (A,Z) (\beta\gamma) = \lambda(A,Z) \; . \; e^{-\frac{((\beta\gamma)-(\beta \gamma)_0^{(Z)}(A))^2}{2\sigma^{(Z)}(A)^2}}
\end{equation}

the velocity function can be simply written as:

\begin{eqnarray}
f^{(A,Z)}(\beta \gamma) & = & G(A,Z) (\beta \gamma) + p(Z) \;.\; G(Z,A-3)(\frac{A}{A-3}.\frac{Z-1}{Z}.\beta \gamma)
\label{eqn:velocity_fit}
\end{eqnarray}

\begin{figure}[t]
\begin{center}
\includegraphics*[width=0.7\textwidth]{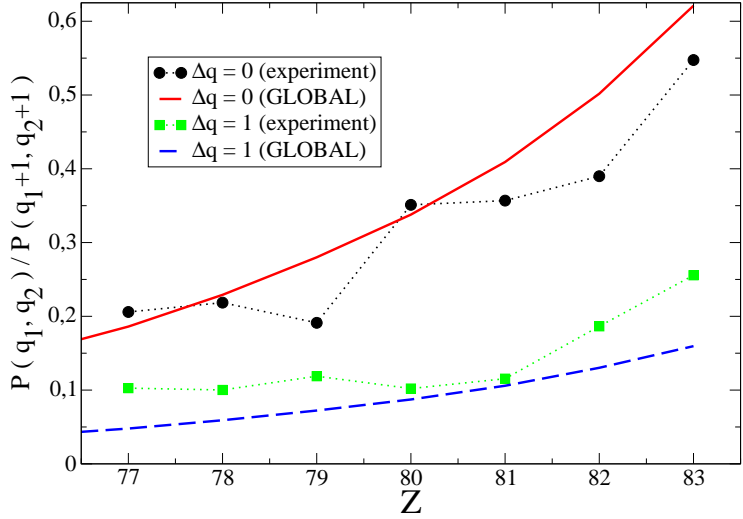}
\caption{\it Comparison between ratios of charge-state probabilities P(0,0)/P(1,1) and P(0,1)/P(1,2) as obtained from the fit of the experimental momentum distributions with the values calculated using the code GLOBAL (see text for details).}
\label{fig:qst_fit}
\end{center}
\end{figure}

The $p(Z)$ coefficient stands for the ratio between the probability of the considered charge state and the probability of the charge state responsible for contamination. It is a free parameter of the fit. The value obtained for each $Z$ can be compared to values calculated with a code dedicated to ionic charge states like GLOBAL~\cite{GLOBAL}. As can be seen on figure~\ref{fig:qst_fit}, in the 500A~MeV Pb+p experiment, the overall agreement was satisfactory.

The experimental production rate $I$ for each isotope, including the contribution from misidentified fragments, is obtained by integration of the $f$ function defined in equation~\ref{eqn:velocity_fit}.

\subsection{Correction of production rates for misidentified fragments}

The contribution from misidentified fragments can be subtracted iteratively:

\begin{equation}
P(q_1,q_2).T(A,Z) = I (A,Z,q_1,q_2) - P(q_1+1,q_2+1).T(A-3,Z)
\end{equation}

As the production cross section of the lightest isotopes decreases exponentially, iteration can be started considering that the net production rate $T$~is 0 for the lightest isotopes. To calculate the charge-state probability $P(q_1+1,q_2+1)$~we used the GLOBAL code. We considered a 10\% uncertainty for this calculation. This leads to a high uncertainty for the few neutron-rich isotopes, for which the correction reaches several tens of percent.

\subsection{Normalization of production rates}

At this step, production rates have been obtained for all the most likely charge-state combinations. In order to reduce the statistical uncertainty, it is appropriate to consider the production rates obtained for all combinations. This differs from high-energy experiments in which it is generally convenient to disregard the non-fully stripped fragments.

\begin{figure}[t]
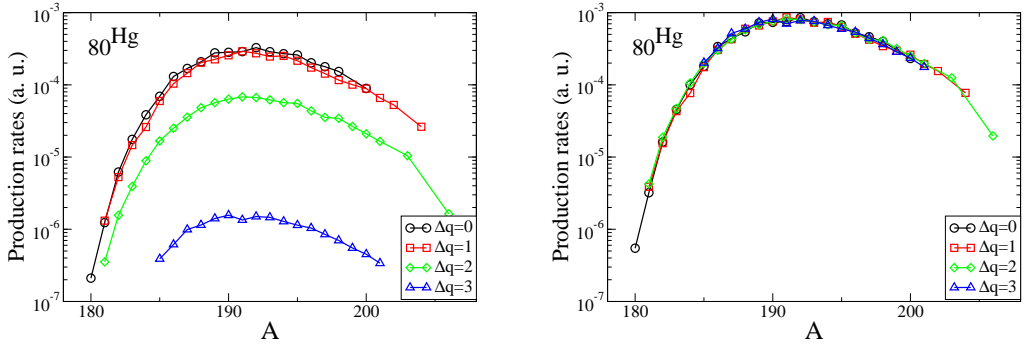

\begin{center}
\includegraphics*[width=0.45\textwidth]{80_raw.eps}
\hspace{0.05\textwidth}
\includegraphics*[width=0.45\textwidth]{80_norm.eps}
\caption{\it Production rates for Hg isotopes obtained for different charge states combinations (left); same data, each distribution being normalized by its respective ionic charge state probability (right).}
\label{fig:qst_renorm}
\end{center}
\end{figure}

The simplest way to normalize all the different production rates consists in using calculated charge-state probabilities, by example using GLOBAL. But it is also possible to fit those probabilities by requiring agreement on the different normalized distributions. This process is illustrated in figure~\ref{fig:qst_renorm} for the Hg isotopes measured in the 500A~MeV experiment. The agreement between the various distributions is very satisfactory. The use of the multiple charge-state combinations has two interests: a reduction of the uncertainty on the production rate and access to an experimental value for the charge-state probabilities.

\begin{figure}[t]
\begin{center}
\includegraphics*[width=0.7\textwidth]{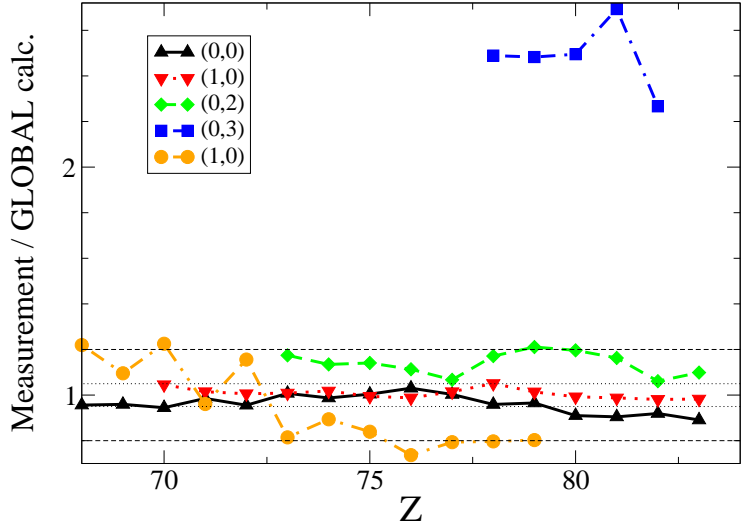}
\caption{\it Comparison of probabilities of charge-state combinations, plotted as the ratio of experimental values (see text) over GLOBAL calculations.}
\label{fig:qst_compare}
\end{center}
\end{figure}

Further check of this method can be achieved by comparing the values obtained for the charge-state probabilities to GLOBAL calculations. This comparison is presented in figure~\ref{fig:qst_compare} for the 500A~MeV experiment. The agreement is better than 10\% for most combinations. Nevertheless a spectacular deficiency of the code is observed for Li-like ions for which production is underestimated by GLOBAL by a factor 2.5.

From this point, the goal of getting the cross sections can be achieved in the same way as in any experiment: one needs to divide the production rates by the number of atoms in the target and by the number of ions from the beam, and correct for the efficiency of the detectors. In the case of fragmentation reactions, multiple reactions play an important role in the production of light isotopes. This is especially true at low energy as the production cross sections fall sharply~\cite{NPA_loa}. Therefore, the contribution of these multiple reactions must be carefully subtracted. A dedicated method has been developed. This is largely self-consistent and uses a limited input from the EPAX parametrization~\cite{EPAX}. This method will be presented in a forthcoming paper~\cite{NPA_loa}.

%*********************************************************************

\section{Conclusions}

%*********************************************************************

It has been demonstrated that a clean $Z$~identification can be obtained for reaction products even when they are not completely stripped by increasing the quantity of gas of the ionization chambers with respect to the quantities used for high-energy experiments.

Conclusions are not so clear regarding mass identification, which requires the charge-state measurement using a thick degrader. According to our calculations, this identification is possible at 500A~MeV, although the expected resolution is extremely close to the acceptable limit. We have shown that during the 500A~MeV Pb+p experiment the obtained resolution was far worse than the calculated value. Using additional simulations we propose that the main reason for this failure is the presence of surface inhomogeneities on the degrader. Therefore, it may be possible that the technique developed for this experiment will be successfully applied in the future, when an improved degrader will be available.

We have demonstrated that it is possible to obtain production rates with a high accuracy, even in the case of a failure of the charge-state measurement. This can be achieved through a systematic evaluation of the contribution of the different charge states and the suppression of the contribution of polluting ones. Calculations by an atomic-physics code (GLOBAL) are needed in order to estimate some corrections, but most of the atomic-physics data can be obtained in a self-consistent way from the data of the experiment. In the case of the 500A~MeV experiment, comparison between GLOBAL calculations and obtained atomic-physics data confirmed the overall accuracy of this code at this energy, with the exception of the estimation of Li-like ions.

If one considers the low-energy limit defined in the early papers dedicated to the FRS, the overall success of the 500A~MeV experiment is clearly a push beyond this limit. Nevertheless, further decrease in energy for experiments in the system Pb+p would create further problems, as further multiplication of the charge states would lead to additional complexity in the analysis.

%*********************************************************************

\appendix

%*********************************************************************

\section{Relation between straggling in energy and magnetic-rigidity width}
\label{chap:brho-e}

An energy loss can be written as a normalized value:

\begin{equation}
\Delta E = A \; (\gamma_1 - \gamma_2)
\end{equation}

Inserting the unit-less version of equation~\ref{eqn:brho}, the previous equation writes:

\begin{equation}
\frac{\Delta E}{A} = \sqrt{1 + ((B\rho)_1.\frac{q}{A})^2} - \sqrt{1 + ((B\rho)_2.\frac{q}{A})^2}
\end{equation}

In section~\ref{chap:uncer} we have shown that the contribution of $(B\rho)_1$~to the total uncertainty is negligible. This leads to the dispersion in $(B\rho)_2$~and the energy loss:

\begin{equation}
\delta ( \Delta E ) = \frac{-((B\rho)_2.\frac{q}{A})^2}{\sqrt{1 + ((B\rho)_2.\frac{q}{A})^2}}  \frac{\delta (B\rho)_2}{(B\rho)_2}
\end{equation}

Dividing by the energy loss and replacing the $B\rho.q/A$~terms by the Lorentz coefficients, one gets:

\begin{equation}
\frac{ \delta ( \Delta E )}{\Delta E} = \beta_2 . \frac {(\beta\gamma)_2}{\gamma_1-\gamma_2} . \frac{\delta (B\rho)_2}{(B\rho)_2}
\end{equation}

One can eventually introduce the difference of magnetic rigidities in the above equation:

\begin{equation}
\label{eqn:brho-e}
\frac{ \delta ( \Delta E )}{\Delta E} = \left( \beta_2 . \frac {(\beta\gamma)_1 - (\beta\gamma)_2}{\gamma_1-\gamma_2} \right) . \frac{(B\rho)_2}{(B\rho)_1 - (B\rho)_2} . \frac{\delta (B\rho)_2}{(B\rho)_2}
\end{equation}

In the energy range considered in this paper (a few hundreds of MeV to one GeV per nucleon) and with the degrader thickness selected as half of the range of the beam, the kinematic factor (left part of the expression in equation~\ref{eqn:brho-e}) is always close to 1 (for example, 0.92 at 500A~MeV or 0.95 at 1A~GeV). This could be expected, as this term may be approximated here by $\beta_2/<\beta>_{1\rightarrow2}$.

%\vspace{1cm}

\end{document}